\documentclass[twocolumn,aps,prb,groupedaddress,showpacs]{revtex4-1}
\usepackage{amsfonts}
\usepackage{amsmath}
\usepackage{amssymb}
\usepackage{txfonts}
\usepackage{pxfonts}
\usepackage{graphicx,bm,units,yfonts}
\usepackage[table]{xcolor}

\begin{document}

\title{The noncommutative Kubo Formula: Applications to Transport in Disordered Topological Insulators with and without Magnetic Fields}
\author{Yu Xue}
\author{Emil Prodan}
\affiliation{Department of Physics, Yeshiva University, New York, NY 10016, USA}
\date{\today }

\begin{abstract}
The non-commutative theory of charge transport in mesoscopic aperiodic systems under magnetic fields, developed by Bellissard, Shulz-Baldes and collaborators in the 90's, is complemented with a practical numerical implementation. The scheme, which is developed within a $C^*$-algebraic framework, enable efficient evaluations of the non-commutative Kubo formula, with errors that vanish exponentially fast in the thermodynamic limit. Applications to a model of a 2-dimensional Quantum spin-Hall insulator are given. The conductivity tensor is mapped as function of Fermi level, disorder strength and temperature and the phase diagram in the plane of Fermi level and disorder strength is quantitatively derived from the transport simulations. Simulations at finite magnetic field strength are also presented. 
\end{abstract}

\maketitle

\section{Introduction.}

Topological insulators represent a new state of matter where the topology stabilizes some extremely robust properties such as the emergence of conducting metallic edge or surface states in spite of the presence of impurities or defects. Topological insulators have been theoretically predicted in 2- and 3-dimensions and now we have several concrete materials that show some of the predicted properties.\cite{HALDANE:1988rh,Kane:2005np,Kane:2005zw,Bernevig:2006hl,Koenig:2007ko,Moore:2007ew,Fu:2007vs,Hsieh:2008vm} For the 3-dimensional topological materials, the surface states have been mapped using angle resolved photoemission spectroscopy and they show an odd number of Dirac cones, in perfect agreement with the theory.\cite{Qi2010,ZHassanRevModPhys2010du,QiRMP2011tu,Hasan2010by} However, the transport measurements have revealed that, so far, all the samples display a metallic character in the bulk. As such, the transport experiments have taken a central stage in the research on topological insulators.

The literature abounds with high quality experimental transport data for topological insulators.\cite{Koenig:2007ko,Koenig:2008so,RothSc2009cu,TaskinPRB2009xu,EtoPRB2010xy,RenPRB2010vu,TaskinPRB2010vi,RenPRB2011vb,RenPRB2011er,TaskinPRL2011bn,RenPRB2012gh,Hor:2009zd,QuScience2010gy,CheckelskyPRL2011bn,JiaPRB2011uv,XiongPhysicaE2012vb,AnalytisPRB2010hv,GusevPRB2011ls,KimPRB2011jf,BrahlekAPL2011vy,SteinbergPRB2011bn,WangPRB2011rt,JeffriesPRB2011fj,AguilarPRL2012sl,SeguraPRB2012rt,CaoPRL2012er,LiuPRL2012bv} There are detailed reports about the behavior of the transport coefficients with the temperature. Magneto-electric measurements are also available, with maps of the conductivity tensor as function of magnetic field strength and electron density, which is controlled by gate voltages or by doping. There are also studies done at finite frequencies and maps of the transport coefficients as functions of films' thickness and disorder strength have been reported. The materials have been progressively tuned, to a point where the contributions to the transport from the bulk and surface are comparable. The available experimental data could be used as a window into the microscopic properties of these materials, if efficient quantitative theoretical analyses could be developed and applied to the real materials. Such analyses will have to include the disorder and the magnetic fields, a task that at first sight may seem extremely difficult. 

We will argue here that actually that is not the case: The disordered systems under magnetic field can be analyzed very much like we analyze the translational invariant systems. Let us take a few lines to explain what we mean by this. For translationally invariant systems, the response and the correlation functions can be conveniently computed in the dual $k$-space. The $k$-space analysis consists of two parts: 
\begin{description}
\item[P1] Derivation of closed-form expressions.
\item[P2] Numerical evaluation of the formulas.
\end{description}

For translational invariant systems, no matter how complex the correlation function, one can easily derive closed-form expressions, which typically involve integrations and derivations of ordinary functions defined over the Brillouin torus. Even though these formulas are formal, in the sense that one still needs a computer to evaluate them, they pretty much fulfill our idea of an analytic solution because the expressions are transparent enough to enable qualitative understanding and, on the quantitative side, the formulas can be numerically evaluated without much effort. 

The analog of the $k$-space calculus for aperiodic systems is the noncommutative Brillouin torus and its noncommutative calculus developed by Bellissard and his collaborators.\cite{BELLISSARD:1994xj,Schulz-Baldes:1998vm,Schulz-Baldes:1998oq} This formalism has been used already in the literature to derive closed-form expressions for the Kubo formula of transport,\cite{BELLISSARD:1994xj,Schulz-Baldes:1998vm,Schulz-Baldes:1998oq} electric polarization and orbital magnetization,\cite{Schulz-BaldesArxiv2012gh} all in the presence of disorder and magnetic fields. We have used the non-commutative calculus to compute topological invariants for disordered topological insulators.\cite{Prodan2010ew,Shulman2010cy,Prodan2011vy,ProdanJPhysA2011xk,XuPRB2012vu} Hastings and Loring have also used the noncommutative setting to derive and to efficiently compute new invariants for disordered topological insulators.\cite{HastingsJMP2010gy,Loring2010vy,HastingsJMP2010gy,Hastings2010by} These applications of the noncommutative calculus pretty much demonstrate that any  response or correlation function, that can be written in closed-form in the $k$-space, can be translated into a closed-form noncommutative expression which incorporates the effect of disorder and magnetic fields. As such, the first part ({\bf P1}) of the analysis is already in place. In a previous work,\cite{ProdanArxiv2012bn} we have made progress on the second part ({\bf P2}) of the analysis, namely on how to quantitatively evaluate the non-commutative formulas. 

For translational invariant systems, the $k$-derivatives are evaluated using finely tuned finite-difference algorithms and the integrals in the $k$-space are evaluated using Riemann sums, both done for various samplings of the Brillouin torus. This leads to approximate results that converge to the exact result as the sampling of the Brillouin torus becomes finer and finer. If the integrands in these formulas are analytic functions of $k$, as it is the case at finite temperatures, the convergence happens exponentially fast. This is an important characteristic of the $k$-space calculus because it enables extremely accurate calculations even with modest computational efforts. Note that in a first-principle calculation, one often deals with hundreds of energy bands so even for $k$-space calculus the efficiency is a big issue. 

Now, when disorder and magnetic fields are present, we have to switch to the noncommutative Brillouin torus and its noncommutative calculus, which are defined in the strict thermodynamic limit. These structures don't have immediate analogs at finite volumes, but as it was elaborated in Ref.~\onlinecite{ProdanArxiv2012bn}, one can define an approximate, finite-dimensional non-commutative Brillouin torus and an approximate non-commutative calculus. These approximate structures can be implemented and manipulated on a computer and, as such, the non-commutative formulas can be numerically evaluated, though in an approximate way. One key result of Ref.~\onlinecite{ProdanArxiv2012bn} is that the approximate results obtained in this way converge exponentially fast to the exact ones as the dimensionality of the non-commutative Brillouin torus is increased. Note that an inverse power law convergence is not sufficient for exploring the typical interesting questions arising for disordered systems. Hence, in some sense, Ref.~\onlinecite{ProdanArxiv2012bn} gave a solution for the second part $({\bf P2})$ of the analysis.

The structure of the present paper is as follows. In the first part we give an introduction of the non-commutative Brillouin torus and its non-commutative calculus, and present a formal derivation of the non-commutative Kubo formula,  closely following Ref.~\onlinecite{Schulz-Baldes:1998vm}. Even though this material has been reviewed with other occasions (see for example Ref.~\onlinecite{BellissardLectNotesPhys2003cy}), an intuitive exposition using a language that is a little more familiar to the condensed matter theorists could be of interest. Furthermore, we thought that the readership will welcome a format where the non-commutative framework, the numerical algorithm and the applications are all presented in one place. 

In the second part we present the approximate finite volume non-commutative Brillouin torus and its non-commutative calculus, together with the emerging approximate Kubo formula. The formula is then broken down to an explicit expression which can be straightforwardly implemented on a computer. 

In the third part we present an application to a model of a disordered 2-dimensional quantum spin-Hall insulator without edges. We map the bulk conductivity tensor as function of Fermi energy ($E_F$), disorder strength ($W$) and temperature. We report several convergence tests. Based on this calculations, we identify the metallic phase and we map the phase diagram of the systems in the $(E_F,W)$ plane. The results give a direct confirmation, via the computation of the conductivity tensor, that strong disorder closes the insulating gap and drives the system into a metallic phase, and then into a topologically trivial phase. When we place the model in the trivial phase, we find that the disorder alone can drive the system into a metallic phase. The phase diagrams are found to be in very good agreement with the ones reported in previous studies. Furthermore, the phase diagram derived from the transport simulations is compared with the phase diagram derived from a level statistics analysis and good agreement is found.

In a second application, we turn the magnetic field on, and we point out a markedly different behavior of the resulting Hofstadter spectra for topological versus non-topological systems. We then map the resistivity tensor as function of Fermi level at fixed magnetic field. We demonstrate that the algorithm is able to resolve the expected Hall plateaus.

\section{Disordered Lattice Models in the Presence of Magnetic Fields}

We will restrict the discussion to 2-dimensional lattice models without edges. This setting is not restrictive because it covers the bulk 2-dimensional Quantum spin-Hall insulators and the films of 3-dimensional insulators (all transport experiments are carried on films). As we shall see, there are many interesting questions arising even for bulk 2-dimensional Quantum spin-Hall insulators, which are investigated in this study.

The Hilbert space of a 2-dimensional lattice model is spanned by functions ${\bm \psi}$ defined over the 2-dimensional lattice $\mathbb{Z}^2$ with values in $\mathbb{C}^D$, where $D$ is the number of quantum states per site. In the clean limit, $D$ is also equal to the number of bands of the model. We will consider a general lattice  Hamiltonian with on-site disorder: 
\begin{equation}\label{MainModel}
\begin{array}{c}
(H_\omega {\bm \psi})({\bm n}) =  \sum\limits_{\bm m} e^{-i \varphi_{{\bm n}{\bm m}}} \hat{t}_{{\bm m}-{\bm n}} {\bm \psi}({\bm m}) 
 +W  \hat{\omega}_{\bm n} {\bm \psi}({\bm n}).
\end{array}
\end{equation}
The phase factor
\begin{equation}
e^{-i \varphi_{{\bm n} {\bm m}}}=e^{-i\pi \phi({\bm n}\wedge  {\bm m})}
\end{equation} encodes the effect of the magnetic field via the Peierls substitution,\cite{PanatiCMP2003rh} where $\phi$ is the magnetic flux per repeating cell, measured in the units of flux quantum $\phi_0=h/e$, and ${\bm n}\wedge  {\bm m}=n_1m_2-m_1n_2$. The $D \times D$ matrix $\hat{t}_{\bm k}$ encodes the hopping amplitudes from a site to its neighboring site situated at the relative distance ${\bm k}$, and $\hat{\omega}_{\bm n}$ is a diagonal $D \times D$ matrix with independent random entries uniformly distributed in the interval $\left [-\frac{1}{2},\frac{1}{2}\right ]$.

The collection of the matrix amplitudes $\{\hat{\omega}_{\bm n}\}_{{\bm n} \in \mathbb{Z}^2}$ can be viewed as a point $\omega$ in the space $\Omega$ defined as: 
\begin{equation}
\begin{array}{c}
\Omega=\varprod \limits _{{\bm n}\in\mathbb{Z}^2}\left [-\frac{1}{2},\frac{1}{2} \right ]^D.
\end{array}
\end{equation}
The space $\Omega$ will be equipped with the probability measure:
\begin{equation}
\begin{array}{c}
d \omega =\prod_{{\bm n}\in \mathbb{Z}^2}\prod_{\alpha=1}^D d\omega_{\bm n}^\alpha.
\end{array}
\end{equation}
The disorder average is then given by the integral $\int_\Omega d\omega\{\ldots\}$. 

There is a natural action of the additive $\mathbb{Z}^2$ discrete group on $\Omega$ given by:
\begin{equation}
\mathfrak{t}_{\bm m}:\Omega \rightarrow \Omega, \ \widehat{(\mathfrak{t}_{\bm m}\omega)}_{\bm n}=\hat{\omega}_{{\bm n}+{\bm m}}.
\end{equation}
The measure $d \omega$ is invariant and ergodic relative to the action of this group.

We would like to point out that accurate lattice models exist for most of the topological materials. They can be derived empirically or from first-principle calculations by following, for example, the methods presented in  Ref.~\onlinecite{LiuPRB2010xf}. These models are believed to be well suited for the transport simulations.

\section{The Noncommutative Framework}
 
For periodic solids, the correlation functions can be cast as closed formulas involving the classic integro-differential calculus over the Brillouin torus. These formulas give the desired answers directly in the thermodynamic limit. Furthermore, by examining the degree of smoothness for the functions entering these explicit expressions, one can easily understand if the formulas are well behaved and how to evaluate them numerically, i.e. how to proceed with the discretization of the classical Brillouin torus and what finite-difference scheme to use for the $k$-derivatives. Similarly, the noncommutative calculus provides an extremely convenient framework to carry the calculations for aperiodic solids and ultimately to derive closed formulas directly in the thermodynamic limit. Furthermore, like in the periodic case, the noncommutative calculus enables one to understand when these formulas are well behaved and how to compute them numerically. 

\subsection{The Abstract Algebra of Observables}

For periodic solids, the integro-differential calculus goes over the algebra of functions defined on the classical Brillouin torus and let us recall a classic result in set topology which says that a topological space can be reconstructed from the commutative algebra of continuous functions defined over that space. As such, the geometric Brilloin torus and the calculus defined over it can be defined in a pure algebraic setting. This is exactly the kind of shift of reasoning that is needed when dealing with aperiodic solids because, when disorder or a magnetic field with irrational flux per repeating cell are present, the geometric Brillouin torus loses its meaning but the commutative algebra of functions defined over the classic Brillouin torus can be easily adapted to the situation. This algebra is replaced by a noncommutative $C^*$-algebra and, in the spirit of what was said above, the resulting noncommutative $C^*$-algebra can be rightfully called the noncommutative Brillouin torus.\cite{BELLISSARD:1994xj}

We now describe this algebra. Its elements are functions  defined on $\Omega \times \mathbb{Z}^2$ with values in the set ${\cal M}_{D\times D}$ of $D \times D$ complex matrices. We will use lower capital letters like $f$, $g$ or $h$ to refer to the elements of the algebra. The addition rule for the algebra is:
\begin{equation}
(f+g)(\omega,{\bm n})=f(\omega,{\bm n})+g(\omega,{\bm n}),
\end{equation}
where the last $``+"$ operation is the ordinary matrix addition. The multiplication rule is:
\begin{equation}
(f*g)(\omega,{\bm n})=\sum_{{\bm m}\in \mathbb{Z}^2} f(\omega,{\bm m})g(\mathfrak{t}^{-1}_{{\bm m}}\omega,{\bm n}-{\bm m})e^{i \pi \phi ({\bm n}\wedge {\bm m})},
\end{equation}
where the product between $f$ and $g$ appearing on the right hand side is the usual matrix multiplication. The algebra has a unit element defined by: ${\bm 1}(\omega,{\bm n})=\delta_{{\bm n},{\bm 0}}$.

The link between the abstract algebra defined above and the operators acting on the quantum states is simple. Each element $f$ generates a covariant family of operators $\pi_\omega f$, which act on a quantum state ${\bm \psi}$ via the following formula:
\begin{equation}
\left((\pi_\omega f){\bm \psi} \right)({\bm n})=\sum\limits_{{\bm m}\in \mathbb{Z}^2} f(\mathfrak{t}^{-1}_{\bm n}\omega, {\bm m}-{\bm n})e^{i \pi \phi ({\bm m}\wedge {\bm n})}{\bm \psi}({\bm m}).
\end{equation}
For example, the Hamiltonian in Eq.~\ref{MainModel} is generated by the element:
\begin{equation}\label{Ham}
h(\omega,{\bm n})= \hat{t}_{\bm n}+W\delta_{{\bm n},{\bm 0}}\hat{\omega}_{\bm n},
\end{equation}
as one can easily verify that: $\pi_\omega h =H_\omega$.

The map $\pi_\omega$ defines a representation of the algebra in the space of operators acting on the quantum states: 
\begin{equation}\label{representation}
\pi_\omega(f* g)=(\pi_\omega f) (\pi_\omega g).
\end{equation} 
The covariant property means:
\begin{equation}
U_{\bm a} (\pi_\omega f) U_{\bm a}^{-1} = \pi_{\mathfrak{t}_{\bm a}\omega}f,
\end{equation}
where $U_{\bm a}$ are the magnetic translations on $\ell^2({\mathbb Z}^d)$:
\begin{equation}
\left(U_{\bm a}{\bm \psi}\right)({\bm n})=e^{ i \pi \phi ({\bm a}\wedge {\bm n})}{\bm \psi}({\bm n}-{\bm a}).
\end{equation}

The algebra defined above can be endowed with the structure of a $C^*$-algebra, a fact that allows one to define a fine spectral theory and a functional calculus, that is, a natural framework to define functions of an element. For this we need a norm with the special property $\|f*g\|\leq \|f\| \|g\|$ and an involution (usually called a *-operation) $f\rightarrow f^*$ such that $\|f*f^*\|=\|f\|^2$.   If $\|\cdot\|'$ denotes the usual norm for operators acting on the quantum states, then the following norm and the star operation:
 \begin{equation}\label{Norm}
\|f\|=\sup_{\omega\in \Omega} \|\pi_\omega f\|', \ \ f^*(\omega,{\bm n})=f(\mathfrak{t}_{-n}\omega,-{\bm n})^\dagger
\end{equation}
have those properties. If we consider only those elements for which the norm in Eq.~\ref{Norm} is finite, then the algebra of observables becomes a $C^*$-algebra which is denoted as ${\cal A}$. It is this algebra that is called the noncommutative Brillouin torus.\cite{BELLISSARD:1994xj} 

\subsection{Spectrum, the resolvent function and the analytic functional calculus}

The eigenvalue problem:
\begin{equation}
H\psi_\lambda = \lambda \psi_\lambda
\end{equation}
 and the functional calculus
 \begin{equation}
 \begin{array}{c}
 \Phi(H)=\sum_\lambda \Phi(\lambda) \ |\psi_\lambda\rangle \langle \psi_\lambda |
\end{array}
 \end{equation}
 for ordinary  self-adjoint matrices can be formulated in a purely algebraic language. In fact, both concepts can be naturally developed and generalized in the abstract setting of $C^*$-algebras, without making any reference to the eigenvectors. A good reference for this topic is the short course in spectral theory by Arveson.\cite{ArvesonBook2002} 

Let us first discuss the notion of the spectrum of an element $f$ from the algebra ${\cal A}$, which generalizes the set of eigenvalues. The points $z$ of the complex plane for which $z-f$ is invertible (in ${\cal A}$) form an open set, called the resolvent set of $f$. The usual notation for this set is $\rho(f)$. The spectrum of the element $f$ is the complement of the resolvent set: $\mathbb{C} - \rho(f)$, and this set is usually denoted by $\sigma(f)$. For any $f$ in a $C^*$-algebra, the spectrum $\sigma(f)$ is a non-empty compact subset of the complex plane. There are two particular classes of elements that we want to mention: 1) the self-adjoint elements: 
\begin{equation}
f^*=f,
\end{equation} 
whose spectra are confined on the real axis, and 2) the unitary elements: 
\begin{equation}
f*f^*=f^**f={\bm 1},
\end{equation}
whose spectra are confined on the unit circle. For example, the Hamiltonian $h$ defined in Eq.~\ref{Ham} is a self-adjoint element while the time evolution generated $h$ is a one-parameter group of unitary elements.

The inverse of $z-f$ will be denoted by $(z-f)^{-1}$. It is an element of ${\cal A}$. When viewed as a function of $z$, $(z-f)^{-1}$ is called the resolvent function of $f$. It is an analytic function of $z$ on the resolvent set $\rho(f)$ with values in ${\cal A}$. Given a function $\Phi(z)$, analytic in a neighborhood of the spectrum of $f$, one can define the function $\Phi$ of $f$ through the formula:
\begin{equation}
\begin{array}{c}
\Phi(f)=\frac{1}{2\pi i}\oint_{\cal C}(z-f)^{-1}dz,
\end{array}
\end{equation}
where ${\cal C}$ is a contour confined to the analytic domain of $\Phi(z)$ and circling the spectrum of $f$. The above formula provides a functional calculus, i.e. a morphism between the algebra of analytic functions in a neighborhood of $\sigma(f)$ and the algebra ${\cal A}$, that is: 
\begin{equation}
(\Phi \cdot \Phi')(f)=\Phi(f)*\Phi'(f),
\end{equation}
for any $\Phi$ and $\Phi'$ two such functions. As such, any algebraic identity for ordinary functions, belonging to the class allowed above, translates automatically into an identity for the same functions but with $z$ replaced by $f$. 

\subsection{The noncommutative integro-differential calculus}

The commutative algebra of ordinary functions defined over the classical Brillouin torus has been replaced by the non-commutative $C^*$-algebra ${\cal A}$. The correlation functions for translational invariant systems are computed via integrals of the form:
\begin{equation}\label{Bhu}
\begin{array}{c}
\int d{\bm k} \ \mathrm{tr}\{\hat{F}({\bm k})\},
\end{array}
\end{equation}
where $\mathrm{tr}$ is the trace over ${\cal M}_{D\times D}$ space. The operation in Eq.~\ref{Bhu} can be seen as a linear functional defined on the space of functions defined over the classical Brillouin torus with values in ${\cal M}_{D\times D}$. This linear functional has two special properties. It is cyclic:
\begin{equation}
\begin{array}{c}
\int d{\bm k} \ \mathrm{tr}\{\hat{F}({\bm k})\hat{G}({\bm k})\}=\int d{\bm k} \ \mathrm{tr}\{\hat{G}({\bm k})\hat{F}({\bm k})\}
\end{array}
\end{equation}
and it is positive:
\begin{equation}
\begin{array}{c}
\int d{\bm k} \ \mathrm{tr}\{\hat{F}({\bm k})^\dagger\hat{F}({\bm k})\}\geq 0.
\end{array}
\end{equation}
These two properties makes the linear functional of Eq.~\ref{Bhu} into a generalized trace.

Now on ${\cal A}$, it is still possible to define a bounded linear functional that replaces the ${\bm k}$-integration. This functional is: 
\begin{equation}\label{NCTrace}
\begin{array}{c}
{\cal T}(f)=\int_\Omega d\omega \ \mathrm{tr}\{f(\omega,0)\},
\end{array}
\end{equation}   
where, like before, $\mathrm{tr}$ is the trace over ${\cal M}_{D\times D}$ space. The linear functional defined in Eq.~\ref{NCTrace} is cyclic and positive:
\begin{equation}
\begin{array}{c}
{\cal T}(f*g) = {\cal T}(g*f), \ \ \  {\cal T}(f*f^*)\geq 0.
\end{array}
\end{equation}
This makes ${\cal T}$ into a generalized trace and ${\cal T}$ replaces the $k$-integration. The existence of a generalized trace is also extremely useful because the algebra ${\cal A}$ can be endowed with a scalar product:
\begin{equation}\label{ScalarProd}
(f,g)={\cal T}(f^*g)
\end{equation}
which makes ${\cal A}$ into a Hilbert space. As we shall see, this is important because the linear maps acting on the space of operators then become linear maps on a Hilbert space and we do know how to manipulate linear maps on Hilbert spaces.

The noncommutative integration defined in Eq.~\ref{NCTrace} is natural in the sense that in the absence of disorder it reduces to the ordinary ${\bm k}$-integration. Furthermore, like the classic integration over the Brillouin torus, the non-commutative integration defined in Eq.~\ref{NCTrace} gives the correlation functions directly in the thermodynamic limit. More precisely, if $F$, $G$, $\ldots$, are the operators acting on the quantum states generated by $f$, $g$, $\ldots$, then:
\begin{equation}\label{TDLimit}
\lim_{\Lambda\rightarrow\infty}\frac{1}{Vol(\Lambda)}\mathrm{Tr}_{\Lambda}\{FG\ldots\} ={\cal T}(f*g*\ldots ),
\end{equation}
where the trace at the left hand side is taken only over the quantum states inside the box $\Lambda$. The identity of Eq.~\ref{TDLimit} not only gives a close formula for the correlation functions for disordered systems under magnetic fields, but also provides a convenient way to determine when a correlation function is well defined in the thermodynamic limit. We will have an entire discussion of this aspect for the conductivity tensor. 

For translationally invariant systems, most if not all the correlation functions of interest are of the form:
\begin{equation}
\begin{array}{c}
\int d{\bm k} \ \mathrm{tr}\{\partial_k^{\alpha_1}\Phi_1(H({\bm k})) \partial_k^{\alpha_2}\Phi_2(H({\bm k})) \ldots \},
\end{array}
\end{equation}
where $\Phi_i(H({\bm k}))$ are some functions of the Hamiltonian and $\partial_k^{\alpha_i}$ are the $k$-derivations (possibly of higher power) on such functions. Thus, the $k$-derivations are important. On the algebra ${\cal A}$, one can define a set of automorphisms to replace the $k$-derivations:
\begin{equation}
\partial_{k_j} \hat{F}({\bm k}) \ \rightarrow (\partial_j f)(\omega,{\bm n})=in_j f(\omega,{\bm n}),
\end{equation}
where $n_j$ is the $j$-th component of ${\bm n}$. The derivations defined above are natural in the sense that in the absence of disorder, they reduce to the ordinary $k$-derivations. Some of the classic rules in integro-differential calculus still apply, such as $\partial_i\partial_j = \partial_j \partial_i$, $\partial_i(f*g)=(\partial_i f)*g + f*(\partial_i g)$ (Leibniz rule) or ${\cal T}(\Phi(h)\partial_i \Phi'(h))=0$. 

We end by pointing out that the representation of the derivation on the quantum states is:
\begin{equation}
\begin{array}{c}
\pi_\omega (\partial_j f)=-i[x_j,\pi_\omega f].
\end{array}
\end{equation}
As such, the element of ${\cal A}$ that generates the charge-current operator is:
\begin{equation}
{\bm j} = -{\bm \nabla}h,
\end{equation}
since:
\begin{equation}
\begin{array}{c}
\pi_\omega {\bm j}=i[{\bm x},\pi_\omega h]=ei[H_\omega,{\bm x}],
\end{array}
\end{equation}
where $e=-1$ is the electron charge.

\section{The noncommutative Kubo formula}

We now have all the rules of calculus and we can proceed with the derivation of the noncommutative Kubo formula. We closely follow Refs.~\onlinecite{BELLISSARD:1994xj} and \onlinecite{Schulz-Baldes:1998vm} and we will present only the main steps of the derivation. The interested reader can consult the ample review article from Ref.~\onlinecite{BellissardLectNotesPhys2003cy}.

\subsection{The coherent time evolution}

The Hamiltonian $h$ itself defines a derivation, i.e. a linear map on the algebra ${\cal A}$ that satisfies the Leibniz rule,\cite{SakaiBook1971gu} through the Liouvillian:
\begin{equation}
\begin{array}{c}
{\cal L}_h[f]=i(h*f-f*h).
\end{array}
\end{equation}
The equation:
\begin{equation}
\begin{array}{c}
\partial_t u(t)=-{\cal L}_h[u(t)], \ u(0)=1,
\end{array}
\end{equation}
defines a one parameter unitary flow $u(t)$ over ${\cal A}$,\cite{SakaiBook1971gu} which implements the time evolution in the Heisenberg picture. Explicitly:
\begin{equation}
u(t)f=e^{-t{\cal L}_h}f=e^{ith}*f* e^{-ith}.
\end{equation}

In the presence of a uniform electric field ${\bm E}$, the Hamiltonian becomes (here we set $e=-1$):
\begin{equation}
h_E=h+{\bm E}\cdot {\bm \nabla}
\end{equation}
and the quantum time evolution is generated as:
\begin{equation}
u_E(t)f=e^{-t{\cal L}_{h_E}}f,
\end{equation} 
with ${\cal L}_{h_E}={\cal L}-{\bm E}{\bm \nabla}$.

\subsection{Scattering processes and the dissipative time evolution}

Decoherence is introduced by random scattering events, such as the electron-phonon scattering. The scattering matrices for various dissipation mechanisms are known explicitly.\cite{BellissardLectNotesPhys2003cy} Here, however, we will proceed at a formal level and assume that the scattering matrix is known and given. Later on, we will adopt the relaxation time approximation.

 If the scattering matrix for a scattering event at time $t$ is denoted by $w_t$, the instantaneous time evolution from initial time $t=0$ to some arbitrary time $t$ takes the form:
\begin{equation}
\begin{array}{c}
u_{w,E}(t)=u_E(t-t_n) w_{t_n}
 u_E(t_n-t_{n-1}) w_{t_{n-1}} \ldots  w_{t_1} u_E(t_1).
\end{array}
\end{equation}
The time sequence $\{t_i\}_{i=1,2,\ldots}$ is assumed to be generated by a Poisson random process with an average frequency $1/\tau$. The scattering matrices $w$ can fluctuate from a scattering event to another. The average of $u_E(t)$ over the collision times and the fluctuations of $w$'s, denoted by $\bar{u}_E(t)$, can be computed explicitly, and the result is:\cite{BELLISSARD:1994xj,Schulz-Baldes:1998vm}
\begin{equation}\label{BarU}
\bar{u}_E(t)=e^{-t(\Gamma+{\cal L}_{h_E})},
\end{equation}
where $\Gamma = (1-\bar{w})/\tau$ with $\bar{w}$ being the average of $w$ over its fluctuations. 

The evolution $\bar{u}_E(t)$ includes the dissipation and is no longer unitary. However, in the absence of an electric field, the dissipative time evolution must leave the thermal equilibrium state un-altered, and this requires that $\Gamma$ and ${\cal L}_h$ commute. 

\subsection{The conductivity tensor}

We assume that the electric field was turned on at $t=0$, when the system was still in its thermal equilibrium state described by the density matrix $\rho_0=\Phi_{FD}(h)$. Here, $\Phi_{FD}$ is the Fermi-Dirac distribution function corresponding to a temperature $T$ and Fermi level $E_F$. After the field was turned on, the density matrix evolves according to: $\rho(t)=u_{w,E}(t)\rho_0$. As such, the instantaneous expected value of the charge current density is:
\begin{equation}\label{CurDen}
{\bm J}_{w,E}(t)={\cal T}\left ({\bm j} * u_{w,E}(t)\rho_0 \right).
\end{equation}
The average over the collision processes leads to:
\begin{equation}\label{CurDen}
\bar{{\bm J}}_E(t)={\cal T}\left ({\bm j} * \bar{u}_{E}(t)\rho_0 \right),
\end{equation}
and the time average:
\begin{equation}
\begin{array}{c}
\langle \bar{{\bm J}}_E \rangle = \lim\limits_{t\rightarrow \infty}\frac{1}{t}\int_0^t dt' \ \bar{{\bm J}}_E(t')
\end{array}
\end{equation}
can be computed as it follows:
\begin{equation}
\begin{array}{c}
\langle \bar{{\bm J}}_E \rangle = \lim\limits_{\delta\rightarrow 0} \delta\int_0^\infty dt \ e^{-\delta t}\bar{{\bm J}}_E(t).
\end{array}
\end{equation}
With the explicit expression of $\bar{{\bm J}}_E(t)$ from Eq.~\ref{CurDen} and the explicit formula for $\bar{u}_E(t)$ from Eq.~\ref{BarU}, the integral over $t$ can be computed explicitly leading to:
\begin{equation}\label{Partial1}
\begin{array}{c}
\langle \bar{{\bm J}}_E \rangle =\lim_{\delta \rightarrow 0} \delta \ {\cal T}\left ({\bm j} * (\delta+\Gamma + {\cal L}_{h_E})^{-1}\rho_0 \right ).
\end{array}
\end{equation}
Since there is no current in the absence of the electric field, we must have:
\begin{equation}
\delta \ {\cal T}\left ({\bm j} * (\delta+\Gamma + {\cal L}_h)^{-1}\rho_0 \right )=0,
\end{equation}
and we can subtract such null term from Eq.~\ref{Partial1}, which then becomes:
\begin{equation}
\begin{array}{c}
\langle \bar{{\bm J}}_E \rangle=
\lim\limits_{\delta \rightarrow 0} \delta \ {\cal T}\left ({\bm j} * (\delta+\Gamma + {\cal L}_{h_E})^{-1} \right . \medskip \\
\left .\circ ({\bm E}{\bm \nabla})\circ(\delta+\Gamma + {\cal L}_h)^{-1}\rho_0 \right ).
\end{array}
\end{equation}
Since ${\cal L}_h \rho_0=0$ and $\Gamma \rho_0=0$ (recall that ${\cal L}_h$ and $\Gamma$ commute), the action of the map seen in the second line of the above equation reduces to just $\delta^{-1}\rho_0$. Hence, we can take the limit $\delta \rightarrow 0$ to finally obtain:
\begin{equation}\label{Current}
\langle \bar{{\bm J}}_E \rangle=-{\cal T}\left (({\bm \nabla}h) * (\Gamma + {\cal L}_{h_E})^{-1}(
{\bm E}{\bm \nabla}\rho_0)\right).
\end{equation}
 The non-linear conductivity tensor can be easily read from above:
\begin{equation}
\begin{array}{c}
\sigma_{ij}({\bm E})=- {\cal T}\left ((\partial_i h) * (\Gamma + {\cal L}_{h_E})^{-1}\partial_j \Phi_{\mathrm{FD}}(h)\right).
\end{array}
\end{equation}
Its limit as $E\rightarrow 0$ gives the linear conductivity tensor: 
 \begin{equation}\label{KuboFormula}
\begin{array}{c}
\sigma_{ij}= -{\cal T}\left ((\partial_i h) * (\Gamma + {\cal L}_h)^{-1}
\partial_j \Phi_{\mathrm{FD}}(h) \right ),
\end{array}
\end{equation} 
and this is the non-commutative Kubo formula. 

In the relaxation time approximation, which we adopt from here on, the map $\Gamma$ is replace by the identity map times a coefficient $1/\tau_{\mathrm{rel}}$. The relaxation time $\tau_{\mathrm{rel}}$ is an empirical parameter, which is assumed as given. The Kubo formula in the relaxation time approximation becomes:
 \begin{equation}\label{KuboFormula}
\begin{array}{c}
\sigma_{ij}= -{\cal T}\left ((\partial_i h) * (1/\tau_{\mathrm{rel}} + {\cal L}_h)^{-1}
\partial_j \Phi_{\mathrm{FD}}(h) \right ).
\end{array}
\end{equation}
This is the expression that will be used in our present transport simulations. For this simplified expression, one can easily show that all the entries are well behaved. In particular, the inverse $(1/\tau_{\mathrm{rel}} + {\cal L}_h)^{-1}$ exists in ${\cal A}$, since the map $i{\cal L}_h$ is a self-adjoint operator in the Hilbert space defined by the scalar product of Eq.~\ref{ScalarProd}, hence the operator $ {\cal L}_h$ does not have spectrum at $-1/\tau_{\mathrm{rel}}$ (or in other words, $-1/\tau_{\mathrm{rel}}$ belongs to its resolvent set hence $1/\tau_{\mathrm{rel}} + {\cal L}_h$ is invertible). Also, the derivation of Fermi-Dirac function of $h$ belongs to the algebra ${\cal A}$ due to its rapid decay to infinity. As such, all the entries in Eq.~\ref{KuboFormula} are well defined hence the non-commutative Kubo formula takes finite values and is numerically stable. We should point out that without the noncommutative formalism, the most one could do was to express the conductivity tensor as a thermodynamic limit (like in Eq.~\ref{TDLimit}) whose existence and stability would have been very difficult to establish. 

\section{An optimal Kubo formula at finite volumes}

The exercise from the previous section, we hope, is fairly convincing in showing that the linear response coefficients, in general, can be expressed as compact and transparent formulas. In this section we discuss how to efficiently evaluate such formulas, which are written directly in the thermodynamic limit and the thermodynamic limit is not accessible to a computer. Once we complete this numerical aspect, we will have a rigorous and practical formalism to investigate the linear response coefficients of disordered systems under magnetic fields. 

\subsection{The $C^*$-algebra over a torus}

We consider a finite square lattice which is wrapped into a torus. The torus is generated by two discrete circles $\mathbb{T}={\cal S}^1_D \times {\cal S}^1_D$. We specifically require that each discrete circle contain $2{\cal N}+1$ points. On this torus we pick an arbitrary point and call it the origin ${\bm o}$. We also introduce a coordinate system on ${\mathbb T}$ by naturally un-rolling the torus onto the points $\{-{\cal N},\ldots,{\cal N}\}^2 \subset \mathbb{Z}^2$, such that the coordinate of the origin is ${\bm 0}$. The coordinates of a point ${\bm p} \in \mathbb{T}$ will be denoted by ${\bm n}_{\bm p}$. The total number of nodes, equal to $(2{\cal N}+1)^2$, will be denoted by $|{\mathbb T}|$.

We now define the group of rotations, which replaces the group of translations on $\mathbb{Z}^2$. Given a point ${\bm p}$ of the discrete torus, we can imagine a succession of rigid rotations of the ${\cal S}^1_D$ circles, that rotate the torus  until the origin ${\bm o}$ reaches the position where the point ${\bm p}$ was located before the rotations. All the other points rotate rigidly with the origin, hence this action defines a map $\mathfrak{r}_{\bm p}$ on the torus. It is clear that $\mathfrak{r}_{\bm p}({\bm o})={\bm p}$, and that $\mathfrak{r}_{\bm p}$ is independent of the specific sequence (which is not unique) of ${\cal S}^1_D$ rotations used to bring the origin at position ${\bm p}$. The rotations satisfy the following commutative group relations:
\begin{equation}
\mathfrak{r}_{\bm q}\circ \mathfrak{r}_{\bm p}=\mathfrak{r}_{\mathfrak{r}_{\bm q}{\bm p}}=\mathfrak{r}_{\bm p}\circ \mathfrak{r}_{\bm q}=\mathfrak{r}_{\mathfrak{r}_{\bm p}{\bm q}}.
\end{equation}

We now define the equivalent $\Omega$ space for the torus, which is denoted by $\Omega_{\cal N}$. Let $\tilde{\omega}=\{\hat{\omega}_{\bm p}\}_{{\bm p}\in \mathbb{T}}$ be a sequence of diagonal $D\times D$ matrices with identical independent random entries uniformly distribution in $[-1/2,1/2]$. Then $\tilde{\omega}$ can be viewed as a point of the probability space $\Omega_{\cal N}=[-1/2,1/2]^{D |\mathbb{T}|}$, which will be endowed with the probability measure: 
\begin{equation}
\begin{array}{c}
d\tilde \omega=\prod_{{\bm p}\in \mathbb{T}}\prod_{\alpha=1}^D d\omega_{\bm p}^\alpha,
\end{array}
\end{equation}
The rotations $\mathfrak{r}$ induce a group of automorphisms on $\Omega_{\cal N}$: 
\begin{equation}
\mathfrak{r}_{\bm q}\tilde \omega = \{\hat{\omega}_{\mathfrak{r}_{\bm q}p}\}_{{\bm p}\in \mathbb{T}},
\end{equation}
whose actions leave the measure $d \tilde \omega$ invariant.

The $C^*$-algebra ${\cal A}_\mathbb{T}$ over the torus is defined as follows. The elements are functions $\tilde{f}$ defined on $\Omega_{\cal N}\times \mathbb{T}$ and taking values in ${\cal M}_{D\times D}$. The law of composition is:
\begin{equation}
\begin{array}{c}
(\tilde{f}*\tilde{g})(\tilde \omega,{\bm p}) \medskip \\
= \sum\limits_{{\bm q}\in \mathbb{T}} \tilde{f}(\tilde \omega,{\bm q})\tilde{g}(\mathfrak{r}_{\bm q}^{-1}\tilde \omega,\mathfrak{r}_{\bm q}^{-1} {\bm p}) e^{i \pi \phi({\bm n}_{\bm p}\wedge {\bm n}_{\bm q})}.
\end{array}
\end{equation}
We need to define a norm and a $*$-operation with the proper characteristics mentioned when we discussed the $C^*$-algebra ${\cal A}$. As before, the norm will be introduced via the operator representations. Each element $\tilde f$ induces an operator acting on $\ell^2(\mathbb{T})\times \mathbb{C}^D$, the square summable sequences defined over the torus with values in $\mathbb{C}^D$: 
\begin{equation}\label{TRep}
\begin{array}{c}
\left ((\tilde{\pi}_{\tilde \omega} \tilde{f})\varphi\right)({\bm p}) 
=\sum\limits_{{\bm q}\in \mathbb{T}} \tilde{f} (\mathfrak{r}_{\bm p}^{-1}\omega,\mathfrak{r}_{\bm p}^{-1}{\bm q}) e^{i \pi \phi({\bm n}_{\bm q}\wedge {\bm n}_{\bm p})}\varphi({\bm q}).
\end{array}
\end{equation}
The map $\tilde \pi_{\tilde \omega}$ provides a representation of the algebra ${\cal A}_{\mathbb{T}}$, i.e.  
\begin{equation}
\tilde{\pi}_{\tilde \omega} (\tilde{f}*\tilde{g})=(\tilde{\pi}_{\tilde \omega} \tilde{f})(\tilde{\pi}_{\tilde \omega} \tilde{g}),
\end{equation}
if and only if the flux $\phi$ (in the units of $\phi_0$) takes a quantized value:
\begin{equation}
\phi = \frac{2}{2{\cal N}+1} \times \mathrm{integer},
\end{equation}
which, from now on, it will always be assumed to be true. We mention that the quantization condition for the flux is in line with Zak's finding that the magnetic translations accept finite representations only if the above quantization is satisfied.\cite{ZakPR1964vy}

In this conditions, one can define the norm:
\begin{equation}\label{TorusNorm}
\|\tilde f \|=\sup_{\tilde \omega \in \Omega_{\cal N}} \|\tilde \pi_{\tilde \omega} \tilde f\|'
\end{equation}
where this time the primed norm denotes the operator norm on $\ell^2(\mathbb{T})\times \mathbb{C}^D$. This norm has the desired property that $\|\tilde f * \tilde g\|\leq \|\tilde f\| \|\tilde g\|$. Furthermore, a $*$-operation can be defined:
\begin{equation}
\begin{array}{c}
\tilde{f}^*(\tilde \omega,{\bm p})=\tilde{f}(\mathfrak{r}_{\bm p}^{-1}\tilde \omega,\mathfrak{r}_{\bm p}^{-1}{\bm o})^\dagger,
\end{array}
\end{equation}
so that the $\tilde{\pi}$ representation and the $*$-operation satisfy the essential relation:
\begin{equation}
\tilde{\pi}_{\tilde \omega} \tilde{f}^* = (\tilde{\pi}_{\tilde \omega} \tilde{f})^\dagger.
\end{equation}
As such, the norm defined in Eq.~\ref{TorusNorm} has the fundamental property:
\begin{equation}
\|\tilde{f}*\tilde{f}^*\|=\| \tilde{f}\|^2,
\end{equation}
which makes ${\cal A}_{\mathbb{T}}$ into a $C^*$-algebra.

\subsection{An approximate integro-differential calculus over the torus}

We now introduce the generalized trace (the integration) over ${\cal A}_{\mathbb{T}}$. It is defined by:
\begin{equation}\label{TorusTrace}
\begin{array}{c}
{\cal T}_{\mathbb{T}}(\tilde{f})= \int_{\Omega_{\cal N}}  d \tilde \omega \ \mathrm{tr}\{\tilde{f}(\tilde \omega,{\bm o})\}.
\end{array}
\end{equation}
This linear functional satisfies all the required properties of a generalized trace, i.e. positivity and cyclicality. The trace can be computed via the equivalent formula:
\begin{equation}\label{AltTorusTrace}
\begin{array}{c}
{\cal T}_{\mathbb{T}}(\tilde{f})= \frac{1}{|\mathbb{T}|}\int_{\Omega_{\cal N}}  d\tilde \omega \ \mathrm{Tr}\{\tilde{\pi}_{\tilde \omega} \tilde{f} \}.
\end{array}
\end{equation}

The differential calculus defined over ${\cal A}$ will not work over the torus because $i {\bm n}_p \tilde f({\bm p})$ does not close continuously at the boundaries of the coordinate system. What we can do is to define an approximate differential calculus, and that is done as follows. Let $\mathfrak x: {\cal S}^1_D \rightarrow \mathbb{R}$ be a continuous function such that ($n_p$ is the coordinate of the point $p\in {\cal S}^1_D$):
\begin{equation}
|\mathfrak x(p)-n_p|=0 \ \mathrm{if} \ |n_p| < {\cal N}/2
\end{equation}
and: 
\begin{equation}
|\mathfrak x(p) |\leq |n_p| \  \mathrm{for \ all} \ p \in {\cal S}^1_D.
\end{equation} 
Then the formula for the approximate derivations is:
\begin{equation}
(\tilde{\partial}_i \tilde{f})({\bm p}) = i\mathfrak x(p_i) \tilde{f}({\bm p}).
\end{equation}
This formula acts like the exact derivation on functions which takes non-zero values only around the origin ${\bm o}$. The elements entering in the Kubo formula, like $\partial_j \Phi_{FD}(h)$, are concentrated near the origin and have a fast exponential decay away from the origin. So the errors introduced by the approximate derivations decay exponentially fast with the size of the torus.

The operator representation of the derivation is as follows. Let ${\cal O}$ be the discrete unit circle in the complex plane defined by the solutions of $z^{2{\cal N}+1}=1$ and let: 
\begin{equation}
\begin{array}{c}
\mathfrak{x}(p) = \sum_{\lambda \in {\cal O}} b_\lambda \lambda^{n_p}
\end{array}
\end{equation}
be the discrete Fourier decomposition of the function $\mathfrak{x}$. Then:
\begin{equation}
\begin{array}{c}
\tilde{\pi}_\omega (\tilde{\partial}_i f) = i\sum_{\lambda\in {\cal O}} b_\lambda \lambda^{-x_i}(\tilde{\pi}_\omega f) \lambda^{x_i},
\end{array}
\end{equation}
where $x_i$ is the $i$-th coordinate operator: $(x_i \varphi)({\bm p})=n_{p_i} \varphi({\bm p})$ for $\phi$ a quantum state over the torus (i.e. in $\ell^2(\mathbb{T})\times \mathbb{C}^D$).

In the numerical simulations, the function $\mathfrak x(p)$ will be chosen in the same way as in Ref.~\onlinecite{Prodan2010ew}, where we computed the Chern number using the non-commutative calculus.

\subsection{The approximate Kubo formula on the torus}

If $h$ is a short-range Hamiltonian in ${\cal A}$, then it also defines a Hamiltonian $\tilde h$ in ${\cal A}_{\mathbb{T}}$:
\begin{equation}
\tilde h(\tilde \omega,{\bm p})=h(\tilde \omega,{\bm n}_{\bm p}),
\end{equation}
provided the torus is large enough. Note that $\tilde \omega\in \Omega_{\cal N}$ is also part of the space $\Omega$. We now can write down the approximate Kubo formula on the torus:
\begin{equation}\label{KuboFormulaTorus}
\begin{array}{c}
\tilde{\sigma}_{i j}=- {\cal T}_{\mathbb{T}}\left ((\tilde{\partial}_i \tilde{h}) * (1/\tau_{\mathrm{rel}}+ {\cal L}_{\tilde{h}})^{-1} \tilde{\partial}_j \Phi_{\mathrm{FD}}(\tilde{h}) \right ).
\end{array}
\end{equation}
The above expression is self-averaging, so the fluctuations with $\tilde \omega$ rapidly disappear as the size of the torus is increased. The most important fact about the above formula is the following error estimate that was established in Ref.~\onlinecite{ProdanArxiv2012bn}:
\begin{equation}\label{Error}
|\tilde \sigma_{ij}-\sigma_{ij}|\leq C_\xi e^{-\frac{1}{2}\xi {\cal N}},
\end{equation}
which tells that Eq.~\ref{KuboFormulaTorus} converges exponentially fast to its thermodynamic limit. Above, $\xi$ is any constant between the bounds $0<\xi< \sinh^{-1}(\kappa/2d\bar{h})$, with $\kappa=\min\{1/2\tau_{\mathrm{rel}},\pi k T\}$ and $\bar{h}$ being the largest hopping amplitude of $h$. The constant $C_\xi$ is fully identifiable and increases with $\xi$.

The formula in Eq.~\ref{KuboFormulaTorus} can be directly evaluated using the standard linear algebra routines, but replacing the matrix operations with the composition rules and the norms corresponding to algebra ${\cal A}_{\mathbb{T}}$, the latter being needed to evaluate the remainders and to advance the iterations (hence the norms introduced above are not just for formalities). In the present work, however, we adopted a more traditional way even though it is probably not very efficient. It is based on the following observations. If $\{\epsilon_a,\varphi_a\}_{i=\overline{1,D|\mathbb{T}|}}$ is the eigensystem for the Hamiltonian $\tilde \pi_{\tilde \omega} \tilde h$:
\begin{equation}
(\tilde \pi_{\tilde \omega} \tilde h) \varphi_a = \epsilon_a \varphi_a,
\end{equation}
then Eq.~\ref{KuboFormulaTorus} can be cast in the following equivalent form:
\begin{equation}\label{PracticalKubo}
\tilde{\sigma}_{ij} = - \frac{1}{|\mathbb{T}|}\sum_{a,b=1}^{D |\mathbb{T}|}\frac{\langle \phi_a|\tilde{\pi}_{\tilde \omega}(\tilde{\partial}\tilde{h})|\phi_b \rangle \langle \phi_b | \tilde{\pi}_{\tilde \omega} (\tilde{\partial} \Phi_{\mathrm{FD}}(\tilde{h})|\phi_a \rangle}{1/\tau + i(\epsilon_a-\epsilon_b)}.
\end{equation}
This expression may look very similar to the usual Kubo formula at finite volume, already used in the previous simulations for disordered systems,\cite{HouaritPPCM1991er,MandalPhysB1998as,RochePRB1999te,SteffenPRB2004bn} but there are a few major differences. Since the derivation $\tilde{\partial}$ has a more involved expression in our case, the derivation of $\Phi_{\mathrm{FD}}(\tilde{h})$ cannot be processed any further. As such, our expression doesn't look like a current-current correlation function.  We can bring our formula to such form if we use a low approximations for $\tilde{\partial}$ but that will destroy the exponential convergence. 

At the end of our theoretical exposition, we want to stress again that the approximate formula Eq.~\ref{KuboFormulaTorus}  and the error estimate Eq.~\ref{Error} should be considered together, because one of them gives a practical way to do computations and the other really tells what is being computed. The $C^*$-algebraic framework has been instrumental for obtaining both results in Ref.~\onlinecite{ProdanArxiv2012bn}. We have also insisted on presenting the $C^*$-formalism because then one could follow a similar philosophy to derive and compute other important linear response coefficients, such as the spin transport coefficients.

\section{Transport simulations for a disordered Quantum spin-Hall Insulator}

\subsection{The model}

We will work with the Bernevig-Hughes-Zhang model,\cite{Bernevig:2006hl} which fits the ``low energy" band theory of the clean HgTe/CdTe wells. In the $k$-space, this effective Hamiltonian takes the form:
\begin{equation}
\begin{array}{c}
H_0({\bm k})=\left ( 
\begin{array}{cc}
h({\bm k}) & \Gamma({\bm k}) \\
\Gamma({\bm k})^\dagger & h^*(-{\bm k})
\end{array}
\right ),
\end{array}
\end{equation}
where $h({\bm k})=\epsilon({\bm k})+{\bm d}({\bm k})\cdot{\bm \sigma}$, with ${\bm \sigma}=(\sigma_x,\sigma_y,\sigma_z)$ encoding the Pauli's matrices, and $\Gamma({\bm k})$ is a $S_z$-nonconserving interaction. The minimal form of the $\Gamma(k)$ matrix is:\cite{Yamakage2010xr}
\begin{equation}
\begin{array}{c}
\Gamma(k)=
i \Lambda \left (
\begin{array}{cc}
\sin k_x - i \sin k_y & 0 \\
0 & \sin k_x +i \sin k_y
\end{array}
\right ).
\end{array}
\end{equation}
The experimentally measured energy bands, in the proximity of the $\Gamma$-point, can be captured by the following expression for ${\bm d}({\bm k})$:
\begin{equation}
{\bm d}= (A \sin k_x,A \sin k_y, M-2B(2-\cos k_x - \cos k_y)).
\end{equation}
The Hamiltonian $H_0$ displays a topological phase if $0$$<$$M/B$$<$$4$ and $4$$<$$M/B$$<$$8$ with the insulating gap closing at $M/B$=0, 4 and 8, and a topologically trivial phase if $M/B$$<$0 or $M/B$$>$8. In our simulations, the parameters will be fixed at: $A=1$, $B=1$, $C=0$, $D=0$, $\Lambda=0.5$, and $\epsilon({\bm k})$ will be also set to zero. The phase diagram of the model, with exactly these same parameter values, has bee investigated in Refs.~\onlinecite{Yamakage2010xr} and \onlinecite{Prodan2011vy}. We will compare the phase diagrams derived from the transport simulations with the phase diagrams from these two references. 

The model can be realized in real space using a square lattice with four quantum states per site. Explicitly, the matrices $\hat{t}_{\bm n}$, to be plugged in Eq.~\ref{MainModel}, are given by:
\begin{equation}
\hat{t}_{(1,0)}=\hat{t}_{(-1,0)}^\dagger=\left (
\begin{array}{cc}
\frac{A}{2i}\sigma_1+B\sigma_3 & \frac{\Lambda}{2} \\
-\frac{\Lambda}{2} & -\frac{A}{2i}\sigma_1+B\sigma_3
\end{array} \right),
\end{equation} 

\begin{equation}
\hat{t}_{(0,1)}=\hat{t}_{(0,-1)}^\dagger=\left (
\begin{array}{cc}
\frac{A}{2i}\sigma_2+B\sigma_3 & \frac{\Lambda}{2i} \sigma_3\\
\frac{\Lambda}{2i}\sigma_3 & -\frac{A}{2i}\sigma_2+B\sigma_3
\end{array} \right)
\end{equation}
and
\begin{equation}
\hat{t}_{(0,0)}=\left (
\begin{array}{cc}
(M-4B)\sigma_3 & 0 \\
0 & (M-4B)\sigma_3
\end{array} \right).
\end{equation}

The onsite disorder is introduced as discussed in the second section, with the restriction that it needs to preserve the time reversal symmetry. For this, the random amplitudes that enter on the diagonal of the matrix $\hat{\omega}_{\bm n}$ (the rest of the entries are zero) must satisfy: $(\hat{\omega}_{\bm n})_{11}=(\hat{\omega}_{\bm n})_{33}$ and $(\hat{\omega}_{\bm n})_{22}=(\hat{\omega}_{\bm n})_{44}$.

\begin{table}[h]\scriptsize
\begin{center}
\small\addtolength{\tabcolsep}{5 pt}
\rowcolors{1}{white}{lightgray}
\begin{tabular}{|c|c|c|c|}
\hline
$E_F$ & $\sigma_{11}$  \tiny{(50x50 lattice)} & $\sigma_{11}$ \tiny{(Exact)} & Error\\
\hline
   0.000&          0.013&          0.013&    0  \\
  -0.368&         0.013&          0.013&    0\%\\ 
  -0.737&         0.013&          0.013&    0\% \\ 
  -1.105&      166.284&     162.006&     2\%\\ 
  -1.474&      390.475&     389.231&     0.3\%\\ 
  -1.842&      428.067 &    429.085&   0.2\%\\ 
  -2.211&      444.573 &    443.920&    0.1\%\\ 
  -2.579&      460.576 &    459.891&    0.1\%\\
  -2.947&      448.714 &    448.550 &    0.1\%\\ 
  -3.316&      421.281&     422.249 &   0.2\%\\ 
  -3.684&      387.195 &    385.555 &    0.4\%\\ 
  -4.053&      341.874&     340.844 &   0.3\%\\ 
  -4.421&      292.086&    289.548 &    0.8\%\\ 
  -4.790&      232.352 &    232.622&    0.1\%\\ 
  -5.158&      171.330 &    170.758 &   0.3\%\\ 
  -5.526&      104.282 &    104.477&    0.2\%\\ 
  -5.895&       34.388 &      34.187&    0.6\%\\ 
\hline
\end{tabular}
\caption{The diagonal conductivity $\sigma_{11}$ computed via Eq.~\ref{PracticalKubo}  for different Fermi energies. The calculations were performed on a $50\times 50$ lattice with the magnetic field and disorder turned off. The table also shows the exact values of $\sigma_{11}$ obtained with the classical Kubo formula and the relative errors ocurring in the first calculation.}
\label{methodcompare}
\end{center}
\end{table}

\subsection{Simple test results}

We show here the most straightforward test calculation, with the disorder and magnetic field turned off. In this case, the conductivity is given by the traditional Kubo formula which can be evaluated in the $k$-space where we can increase the sampling of the Brillouin torus until the computation becomes virtually exact. At the same time, we can still evaluate the conductivity via Eq.~\ref{PracticalKubo}. A comparison between the two results will provide the test.  

Table \ref{methodcompare}  lists $\sigma_{11}$ as function of Fermi energy, with the noncommutative Kubo formula of Eq.~\ref{PracticalKubo} evaluated on a $50\times50$ lattice. The table also lists the virtually exact $\sigma_{11}$ values obtained with the traditional Kubo formula in $k$-space. As the table shows, all the errors are less than one percent, except in one case where we see a 2\% error (due to a spike in the density of states at that energy). Based on this table we expect well converged results when the calculations are performed on $50 \times 50$. When disorder is present, the convergence will be double-checked by comparing the outputs from calculations performed on lattices of increasing sizes, typically $30 \times 30$, $40 \times 40$ and $50 \times 50$. As we shall see, the convergence of the calculations is also confirmed by those tests.
 
\subsection{The diagonal conductivity as function of disorder and Fermi level}

\begin{figure*}
\center
  \includegraphics[width=16cm]{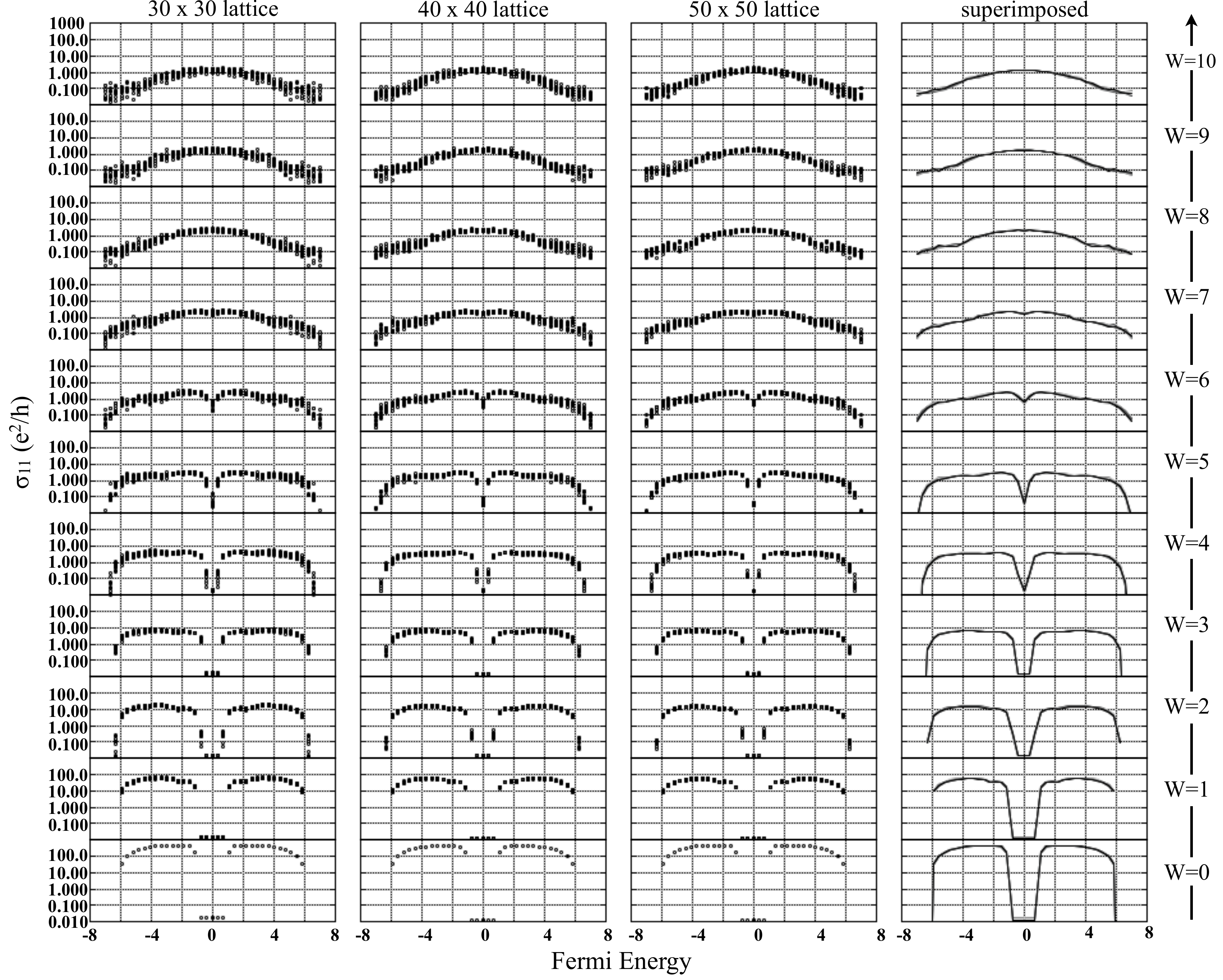}\\
  \caption{The diagonal conductivity for the topological case ($M=6$) as function of Fermi level, lattice size and disorder strength. For each Fermi level, the noncommutative Kubo formula was evaluated for 10 random disorder configurations, with $kT=1/\tau_{\mathrm{rel}}=0.01$. The last column shows the averages over the disorder configurations. The averages corresponding to the three lattice sizes overlap each other almost perfectly indicating a good convergence of the calculations with the lattice size.}
 \label{ConductivityDisorderTI}
\end{figure*} 

Fig.~\ref{ConductivityDisorderTI} reports the calculated $\sigma_{11}$ for the topological case $M=6$ (the off-diagonal components of $\sigma$ are zero because of time reversal symmetry). The diagonal conductivity is plotted as function of Fermi energy at various disorder strengths.  The first three columns show the results obtained by evaluating Eq.~\ref{PracticalKubo} on $30 \times 30$, $40 \times 40$ and $50 \times 50$ lattices. The temperature and the relaxation time were fixed at: $kT=1/\tau_{\mathrm{\mathrm{rel}}}=0.01$. The calculations were repeated for 10 random disorder configurations and the results are all shown in the first three columns of Fig.~\ref{ConductivityDisorderTI} without averaging (hence the fuzziness displayed in those graphs). The (vertical) spread of $\sigma_{11}$ due to the changing of the disorder configuration tells us about the degree of self-averaging in these calculations. Clearly the spread decreases, and as such, the self-averaging is improved as the lattice size is increased. But the most important fact about these data, are the averages of $\sigma_{11}$ over the disorder configurations shown in the 4-th column of Fig.~\ref{ConductivityDisorderTI}. There are minor or no visible variations between the averages obtained on different lattice sizes, a fact that indicates a very well converged simulation. With that, we fulfilled the main goal of this work, namely, to demonstrate the efficiency and effectiveness of the non-commutative Kubo formula.

We now discuss the data. In Fig.~\ref{ConductivityDisorderTI}, one can see that, if the Fermi level is located in certain energy regions, $\sigma_{11}$ rapidly decreases as the disorder strength is increased.  But one can also see well defined energy regions where $\sigma_{11}$ saturates and stays at a certain appreciable value even at large disorder strengths. As we shall see from the temperature-dependence analysis, the latter spectral regions are metallic in character, while the former ones are insulating.

Fig.~\ref{ConductivityDisorderTI} also displays a pattern that is very specific to topological models. The energy regions where $\sigma_{11}$ is maximum are seen to drift towards each other and ultimately to merge together as the disorder strength is increased. This phenomenon is called the levitation and annihilation of the conducting states. Its origin is connected to the fact that the conducting states carry a non-trivial topological $\mathbb{Z}_2$ invariant. Since this invariant is robust against strong disorder, the conducting states cannot suddenly disappear and, instead, neighbouring conducting states levitate towards each other and annihilate their $\mathbb{Z}_2$ invariants. From there on, the states can localize, as it inherently happens in any system if the disorder strength is large enough.\cite{Aizenmann1993uf}  This phenomenon was widely discussed in the context of Integer Quantum Hall effect,\cite{HalperinPRB1982er,LaughlinPRL1984mc,RochePRB1999te,YangPRB1999de,KivelsonPRB1992bv} and it was first observed in the context of Quantum spin-Hall Insulators in Ref.~\onlinecite{Onoda:2007xo} and in Chern insulators in Ref.~\onlinecite{Prodan2010ew}.

\begin{figure}
\center
  \includegraphics[width=8.6cm]{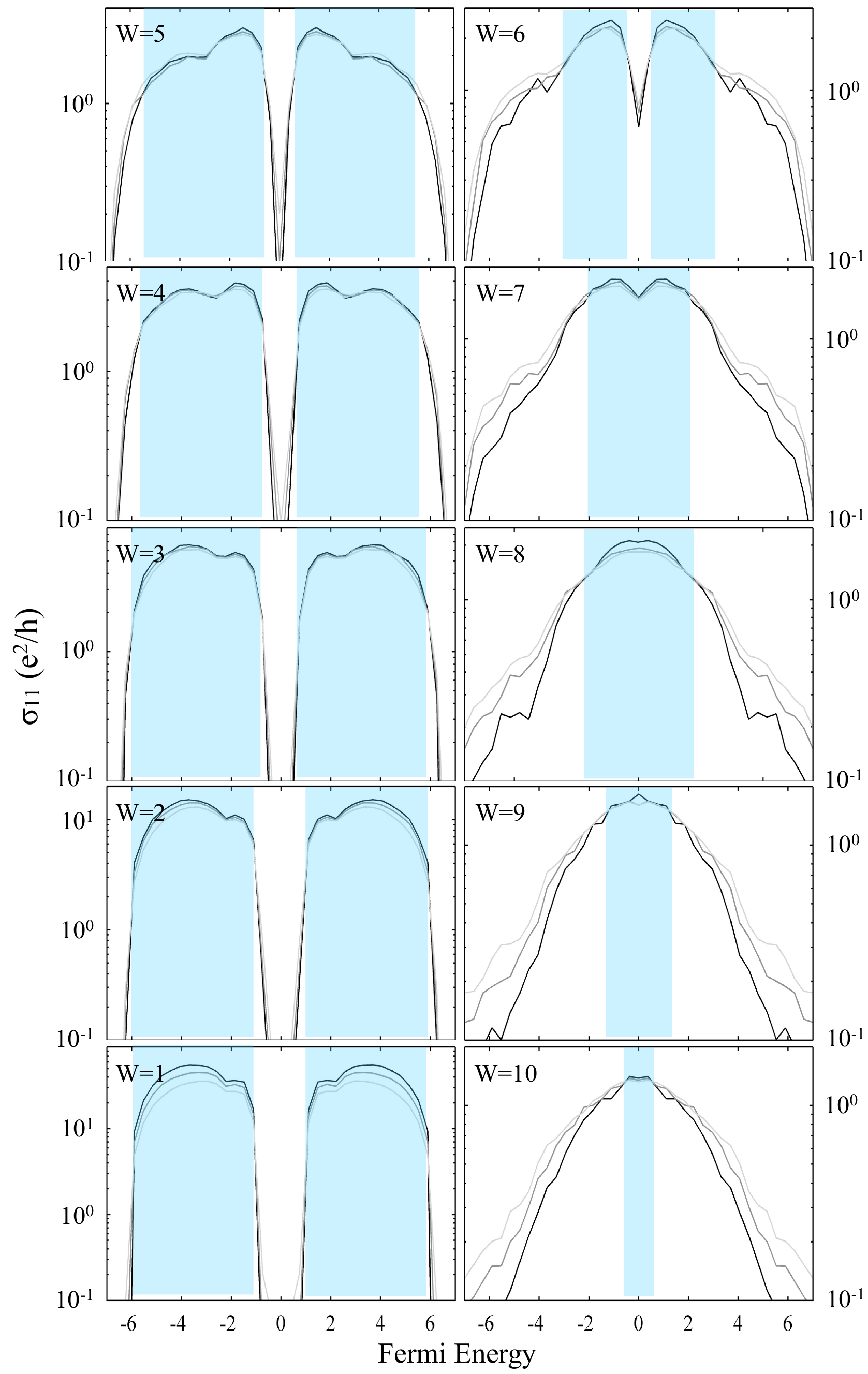}\\
  \caption{The diagonal conductivity for the topological case ($M=6$) as function of Fermi energy, disorder strength and temperature. An average over 10 disorder configurations was used. Each panel displays three curves, corresponding to $kT=1/\tau_{\mathrm{rel}}=0.01$ (black), $kT=1/\tau_{\mathrm{rel}}=0.025$ (gray) and $kT=1/\tau_{\mathrm{rel}}=0.05$ (light gray). The shaded regions indicate the Fermi energies where $\sigma_{11}$ increases when the temperature is reduced, i.e. where the model displays a metallic behavior.}
 \label{CondTempDepTop}
\end{figure} 

\subsection{Temperature dependence and the phase diagram}

The relaxation time $\tau_{\mathrm{rel}}$ behaves as $\tau_{\mathrm{rel}} \sim 1/(kT)^\alpha$ ($\alpha>0$) in the limit of low temperatures, where the exponent $\alpha$ depends on the dominant dissipation mechanism ($\alpha=5$ if the dissipation is through phonons). As such, the temperature dependence of $\sigma_{11}$ comes from the Fermi-Dirac statistics and from $\tau_{\mathrm{rel}}$. The behavior of $\sigma_{11}$ as function of $T$ in the asymptotic limit $T \rightarrow 0$ can provide an accurate picture of the nature of the energy spectrum and the transport characteristics of the system.

The spectral, fractal and diffusion exponents and their inter-relations and relevance to transport in aperiodic solids were introduced in Ref.~\onlinecite{Schulz-Baldes:1998vm}. In particular, the diffusion exponent $\beta_{\mathrm{diff}}$ at the Fermi level was defined through the mean square of the displacement operator in the asymptotic limit of large time intervals:
\begin{equation}
\delta x^2(t)=\frac{1}{t}\int_0^t dt' \ {\cal T}\left (({\bm x}(t')-{\bm x})^2\pi_{E_F}\right ) \sim t^{2\beta_{\mathrm{diff}}}.
\end{equation}
Here ${\bm x}$ is the position operator, ${\bm x}(t)$ is its time evolution and $\pi_{F_F}$ is the projector onto the energy spectrum below $E_F$. The following behavior of $\sigma_{11}$ at low temperatures was proved in Ref.~\onlinecite{Schulz-Baldes:1998vm}:
\begin{equation}\label{Drude}
\sigma_{11}\sim T^{\alpha(1-2\beta_{\mathrm{diff}})}.
\end{equation}
This relation is the Drude formula for aperiodic solids.

The diffusion exponent takes values between 0 and 1. Its precise value can give information about the nature of energy spectrum. In general, the energy spectrum can be absolute-continuous, singular-continuous (fractal) or pure point (localized). The transport is called ballistic if $\beta_{\mathrm{diff}}=1$ and that requires the spectrum around $E_F$ to be absolute continuous. The reverse statement is true in one dimension, but in higher dimensions there are models with absolute continuous spectrum but with non-ballistic (diffusive) transport.\cite{Bellissard:2000hk} The transport is called diffusive if $0<\beta_{\mathrm{diff}}<1$ and occurs when the energy spectrum around $E_F$ is singular-continuous. Absence of diffusion $\beta_{\mathrm{diff}}=0$ occurs when the spectrum is pure point (localized). A detailed discussion of $\beta_{\mathrm{diff}}$ for different quantum models, together with the physical implications, was given in Refs.~\onlinecite{Bellissard:2000lj} and \onlinecite{BellissardLectNotesPhys2003cy}. We want to point out that the exponent in Eq.~\ref{Drude} was experimentally evaluated in Ref.~\onlinecite{CheckelskyPRL2011bn} for a film of Bi$_2$Se$_3$. We believe this exponent can be extracted from many other experimental data.

While we do have the technology to compute the diffusion exponent for our topological model, here we address a coarser problem, namely, we determine when the system behaves as a conductor: $\sigma_{11} \nearrow \infty$ as $T \searrow 0$, or as an insulator: $\sigma_{11} \searrow 0$ as $T \searrow 0$ . This is something we can do at the grand-scale of Fig.~\ref{ConductivityDisorderTI}, which then will allow us to draw the phase diagram of the system. A study on the $\beta_{\mathrm{diff}}$ itself will be reported elsewhere.

\begin{figure*}
\center
  \includegraphics[width=16cm]{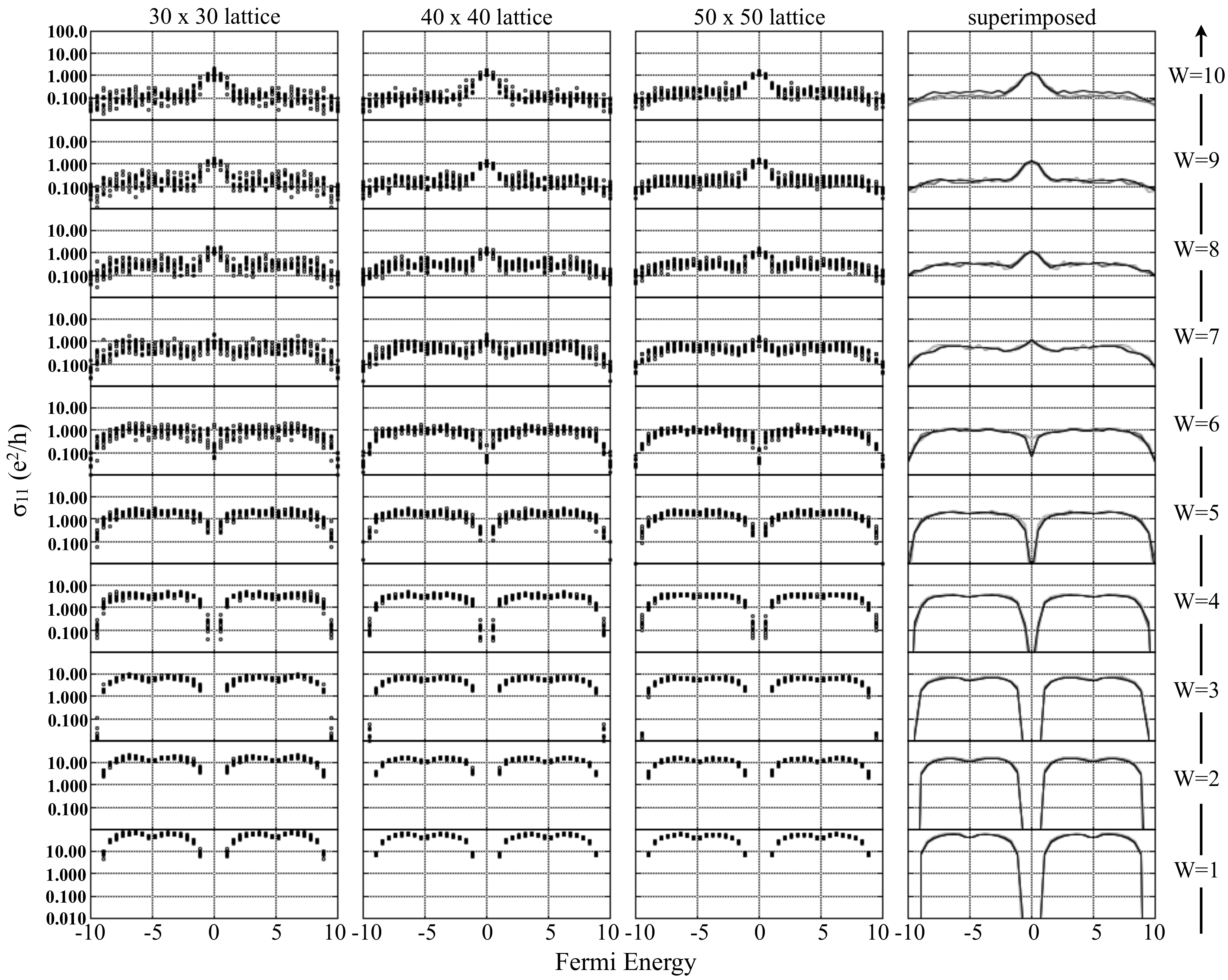}\\
  \caption{The diagonal conductivity for the trivial case ($M=9$) as function of Fermi level, lattice size and disorder strength. For each Fermi level, the noncommutative Kubo formula was evaluated for 10 random disorder configurations, with $kT=1/\tau_{\mathrm{rel}}=0.01$. The last column shows the averages over the disorder configurations. The averages corresponding to the three lattice sizes overlap each other almost perfectly indicating a good convergence of the calculations with the lattice size (maybe with the exception of the last panel).}
 \label{ConductivityDisorderTrivial}
\end{figure*} 

For this we repeated the calculations from Fig.~\ref{ConductivityDisorderTI} for: $kT=1/\tau_{\mathrm{rel}}=0.025$ and $kT=1/\tau_{\mathrm{rel}}=0.05$. We thus assume an exponent $\alpha=1$ but note that the conducting or insulating character is independent of this exponent. The theory predicts a much faster convergence for these cases towards the thermodynamic limit, so we restricted these calculations to only a $40\times 40$ lattice. The results are reported in Fig.~\ref{CondTempDepTop}. According to the previous discussion, in these plots we should see conducting energy regions where the values of $\sigma_{11}$ increase as $T$ is lowered and insulating energy regions where $\sigma_{11}$ decreases as $T$ is lowered. This is indeed consistently observed in Fig.~\ref{CondTempDepTop} for all three curves corresponding to different temperatures. For example, all three curves intersect each other at more or less the same point, so the transition point between the conducting and insulating regions can be identified. As expected, the conducting energy regions occur in the middle of the bands while the insulating regions occur near the edges. In agreement with the levitation and annihilation picture, the conducting regions  are seen to drift towards each other until they merge, at about $W=7$, at which point they rapidly diminish as $W$ is being further increased.

\subsection{Disorder induced conducting states}

The model used in our simulations can enter the Quantum spin-Hall topological phase upon increasing the disorder, even if we start from the trivial phase at $W=0$.\cite{LiPRL2009xi,GrothPRL2009xi,JiangPRB2009sf,Yamakage2010xr,Prodan2011vy} That is due to a strong disorder-induced deformation of the Quantum spin-Hall phase boundary.\cite{Prodan2011vy} We should mention that this phenomenon can disappear if other types of disorder are used.\cite{SongPRB2012gf} The phase diagram of the model with onsite disorder was computed by various methods such as by mapping the conductance of the edge states,\cite{LiPRL2009xi,GrothPRL2009xi,JiangPRB2009sf} the Lyapunov exponent,\cite{Yamakage2010xr} or the spin-Chern number.\cite{Prodan2011vy}

With the Rashba interaction turned on, the model belongs to the symplectic class so the topological and trivial phases are separated by a metallic phase.\cite{Yamakage2010xr,Prodan2011vy,ProdanJPhysA2011xk} As such, we can use our transport simulations to see if indeed there are metallic states induced at large disorder, when such states are absent at weak disorder. In principle, we can use the transport simulations to map the whole metallic phase between the trivial and the topological insulating phases, in the 3-dimensional space of $E_F$, $W$ and $M$, and that will be the fourth way to compute the phase diagram of the model. The fifth way will be to use the level statistics analysis as it was done for other models of disordered topological insulators.\cite{Prodan2010ew,ProdanJPhysA2011xk,XuPRB2012vu,LeungPRB2012vb} 

\begin{figure}
\center
  \includegraphics[width=8.6cm]{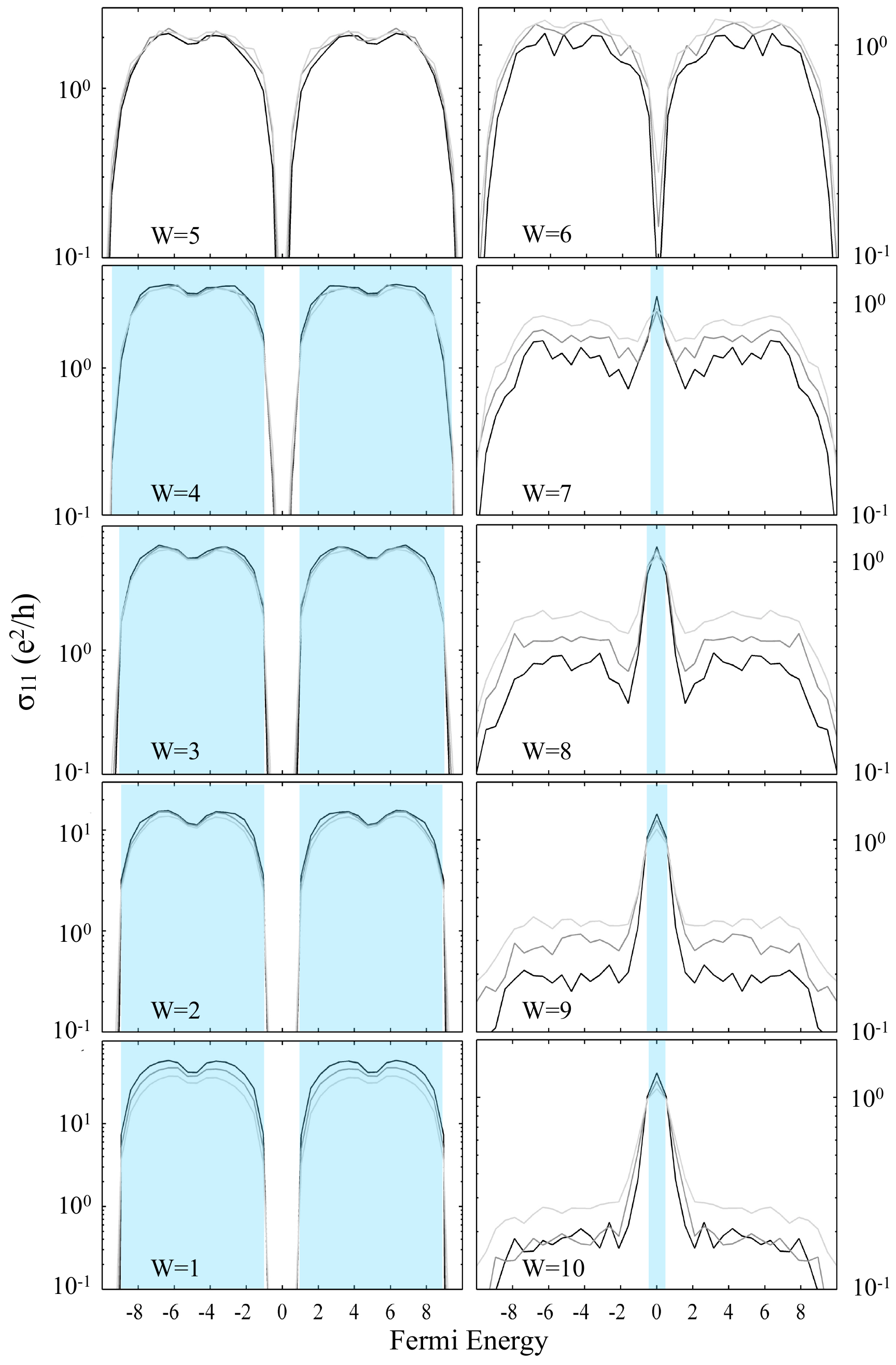}\\
  \caption{The diagonal conductivity for the trivial case ($M=9$) as function of Fermi energy, disorder strength and temperature. An average over 10 disorder configurations was used. Each panel displays three curves, corresponding to $kT=1/\tau_{\mathrm{rel}}=0.01$ (black), $kT=1/\tau_{\mathrm{rel}}=0.025$ (gray) and $kT=1/\tau_{\mathrm{rel}}=0.05$ (light gray). The shaded regions indicate the Fermi energies where $\sigma_{11}$ increases when the temperature is reduced, i.e. where the model displays a metallic behavior.}
 \label{CondTempDepTrivial}
\end{figure}

However, in this section we will compute just another slice of the phase diagram, corresponding to $M=9$ which is in the trivial part of phase diagram at $W=0$. We have repeated the simulations reported in Fig.~\ref{ConductivityDisorderTI} (with $M$ set at the new value) and the results are reported in Fig.~\ref{ConductivityDisorderTrivial}. Here, we see again a reduction in the spread due to the disorder of $\sigma_{11}$ as the size of the lattice is increased, and a good overlap of the disorder-averages for the different lattice sizes. While for the most part of the spectrum the conductivity rapidly decay with the increase of $W$, one can see in the middle of the spectrum and starting from $W=7$ a sudden increase of $\sigma_{11}$ until it reaches the same value as in Fig.~\ref{ConductivityDisorderTI}. To demonstrate that the system enters a metallic phase, we have repeated the simulations from Fig.~\ref{CondTempDepTop} giving $M$ the new value. The results are reported in Fig.~\ref{CondTempDepTrivial} and they indeed confirm that, starting from $W=7$, there is a region where $\sigma_{11}$ increases as the temperature is lowered, hence those states have metallic character.

We will like to point out that our conclusions based on the transport calculations are in quantitative agreement with the phase diagrams computed in Refs.~\onlinecite{Yamakage2010xr} and \onlinecite{Prodan2011vy}. For example, our data in Fig.~\ref{ConductivityDisorderTI} show that at $W=7$ the topological phase is completely gone, and only the metallic and the trivial phases are still present. This is precisely what was predicted in these two mentioned studies. Furthermore, it was predicted that if $M=9$, then the model will enter the metallic phase at about $W=7$ which is exactly what we observe in our transport simulations.

\section{Comparison with the level statistics}

\subsection{The topological case}

\begin{figure}
\center
  \includegraphics[width=7cm]{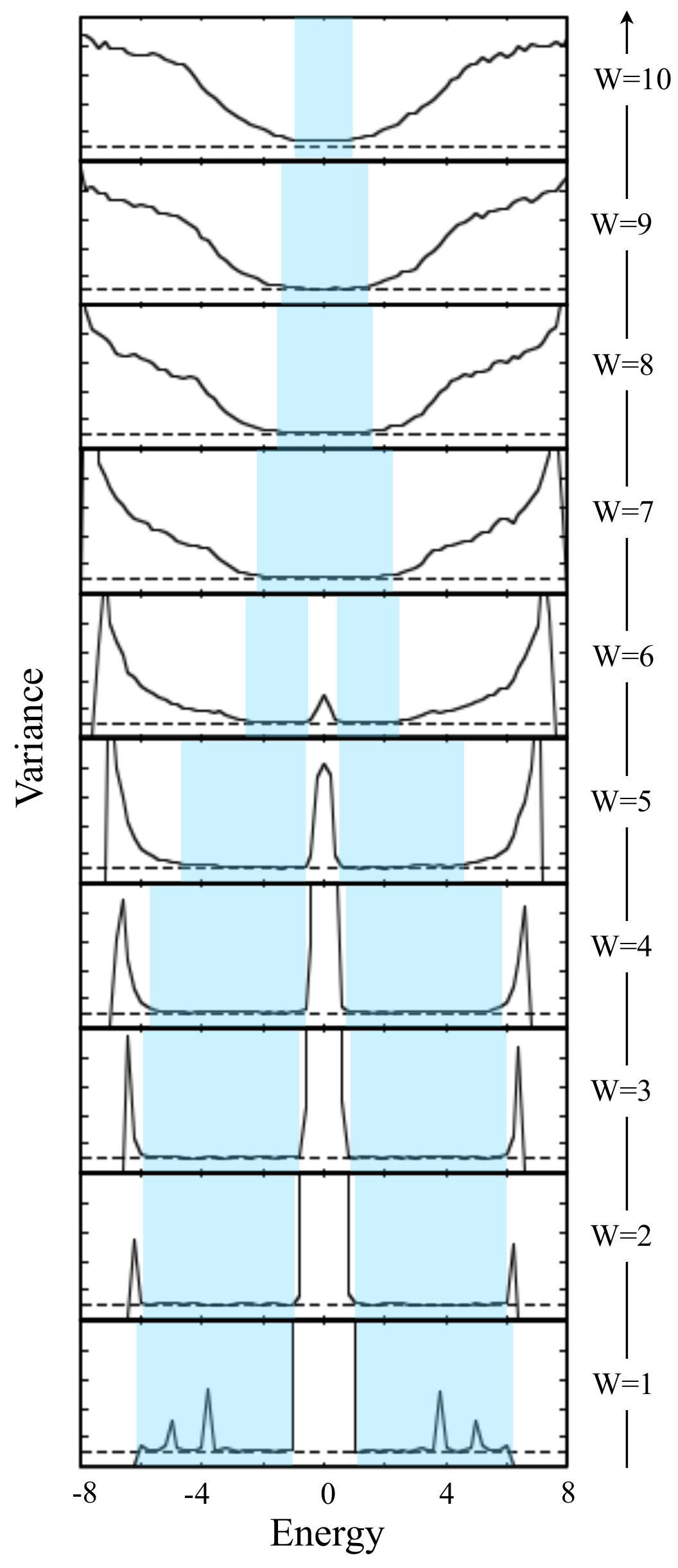}\\
  \caption{The variance of the ensembles of level spacings for the topological case ($M=6$), recorded at various energies as function of disorder strength. A total of 500 disorder configurations were used in these simulations and, for each disorder configuration and energy $E$, 13 level spacings were collected from the immediate vicinity of $E$. As such, the ensembles contain 6,500 level spacings. The size of the lattice for these simulations was $40 \times 40$.}
 \label{LevelStatisticsTop}
\end{figure}

We performed a level statistics analysis in precisely the same manner as in our previous studies.\cite{Prodan2010ew,ProdanJPhysA2011xk,XuPRB2012vu,LeungPRB2012vb} The exact procedure for the level statistics analysis was described in detail in these publications. Fig.~\ref{LevelStatisticsTop} reports the variance of the distribution of the level spacings collected from small energy windows centered at various energies while the random potential was updated 500 times. The mass term was fixed at $M=6$. It is a well established fact that,\cite{EfetovBook1997vn} if the localization length of the system is smaller than the simulation box, then the level spacings follow a Poisson distribution which has variance equal to 1, and if the localization length is comparable or larger than the simulation box, then the level spacings follow the Wigner surmise for symplectic Gaussian ensembles which has a variance of 0.104. As such, the Fig.~\ref{LevelStatisticsTop} allows us to identify the spectral regions with very large or infinite localization lengths, which coincide with the energy intervals where the variance is close to expected value of 0.104. The level statistics analysis was performed on a 40$\times$40 lattice.

The energy regions harboring the extended states have been shaded in Fig.~\ref{LevelStatisticsTop} for a better visualization. The phenomenon of levitation and annihilation is clearly displayed there and the emerging phase diagram is in good quantitative agreement with Fig.~\ref{CondTempDepTop}. Strictly speaking, the metallic region identified in Fig.~\ref{CondTempDepTop} is strictly contained (thus not equal) inside  the region of extended states identified in Fig.~\ref{LevelStatisticsTop}. That is because the metallic phase in Fig.~\ref{CondTempDepTop} contains all the states with $\beta_{\mathrm{diff}}>0.5$ while the phase of extended states in Fig.~\ref{LevelStatisticsTop} contains all the states with $\beta_{\mathrm{diff}}>0$.

\subsection{The disorder induced  conducting states}

\begin{figure}
\center
  \includegraphics[width=7cm]{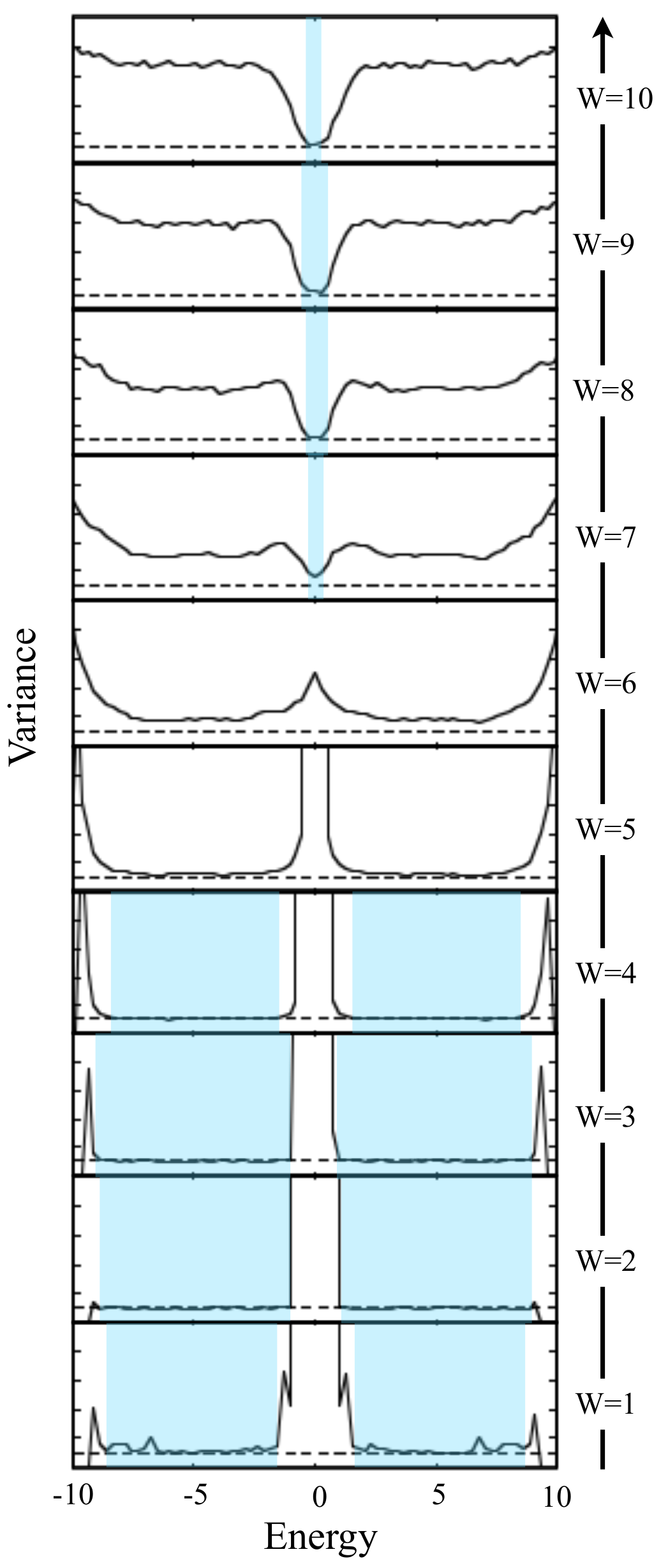}\\
  \caption{Same as Fig.~\ref{LevelStatisticsTop} for the trivial case ($M=9$).}
 \label{LevelStatisticsTrivial}
\end{figure}

We have repeated the level statistics analysis for $M=9$ and the results are reported in Fig.~\ref{LevelStatisticsTrivial}. We have again shaded the energy regions that harbor extended states and, sure enough, we observe the emergence of the extended states that were previously seen in our transport simulations reported in Fig.~\ref{CondTempDepTrivial}. Besides these interesting conducting states, one can also see the extended states that originate directly from the bands of the clean system. We should probably mention that in a 2-dimensional symplectic model, unlike the unitary Anderson model, the extended states can survive moderate disorder. But there is a distinct difference in the way these extended states and the extended states in Fig.~\ref{LevelStatisticsTop} behave as the disorder is increased. In the former case, the levitation and annihilation is absent and the energy domains harboring the extended states simply contract and vanish.

\section{Transport simulations for disordered Quantum spin-Hall Insulators under magnetic fields} 

The time-reversal symmetry is essential for the topological properties of the Quantum spin-Hall Insulators. Nevertheless, the transport experiments in the presence of a magnetic field, which breaks the time-reversal symmetry, are extremely useful for understanding the microscopic electronic structure of the samples. In a typical experiment, the transport coefficients are mapped as functions of the magnetic field strength and of a gate potential. The Hall resistivity measurements at weak magnetic fields provide an accurate map of the electron density as function of the gate potential, and typically this function has a simple linear shape. As such, the dependance of the transport coefficients on the magnetic field, electron density and temperature can be accurately determined. 

\begin{figure}
\center
  \includegraphics[width=8.6 cm]{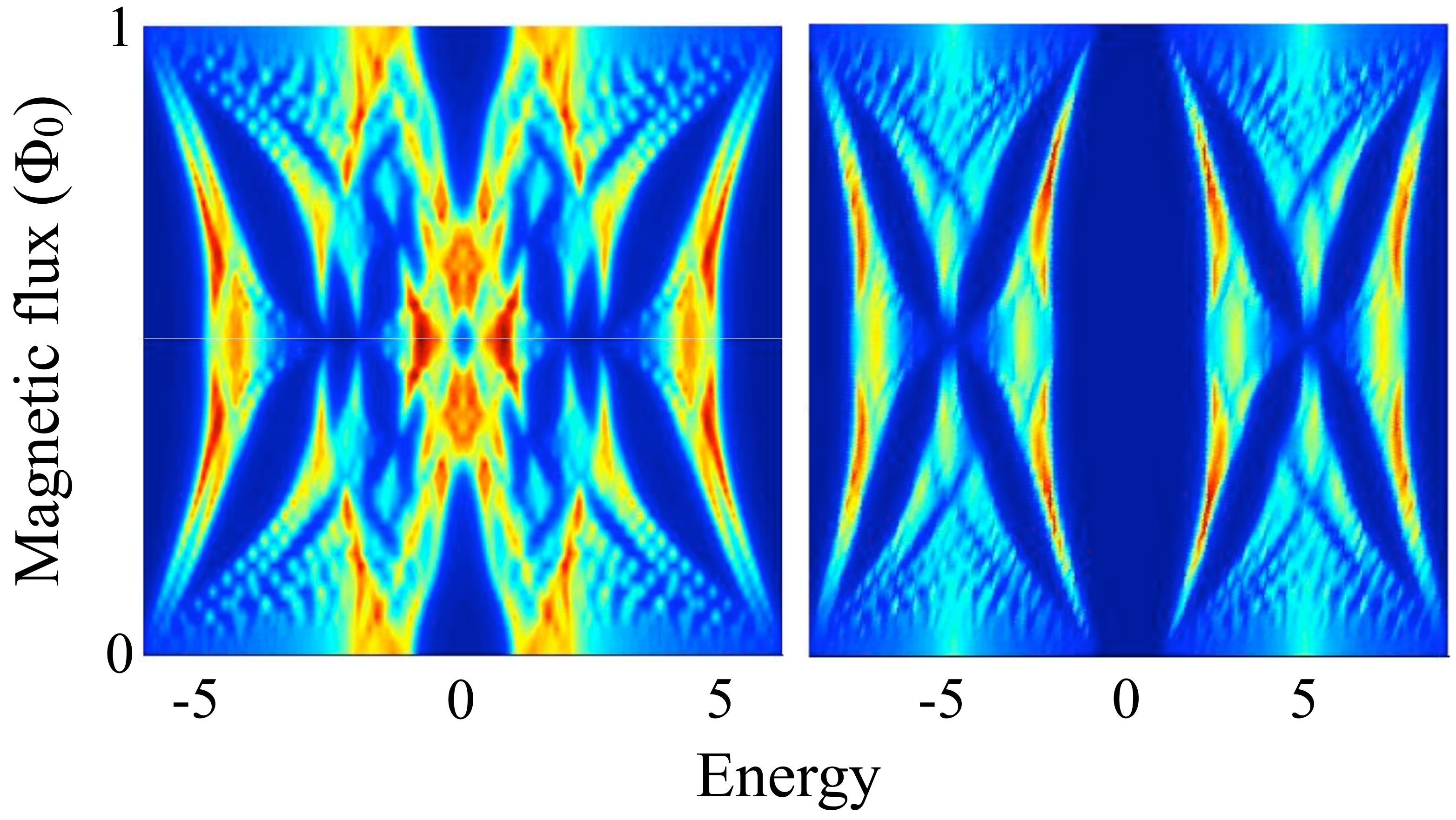}\\
  \caption{Evolution of the density of states (DOS) for the topological case $M=6$ (left) and trivial case $M=9$ (right) with the variation of the magnetic field. The lattice size for this simulations was 100$\times$100 and $W=0$.}
 \label{BDOSW0}
\end{figure}

A typical transport measurement on HgTe Quantum spin-Hall insulators or thin films of 3-dimensional topological insulators shows the following qualitative features: Initially, the Hall resistivity increases (decrease) linearly for n-type (p-type) carrier concentration as the magnetic field is turned on. Hall plateaus start to appear at some point and they become fairly  wide and well defined at larger field strengths, typically of a few Teslas. The diagonal resistivity starts flat as the magnetic field is strengthened, but at some point it starts to develop well defined oscillations of increasing amplitude, the Shubnikov-de Haas oscillations. At strong magnetic fields, where the Hall plateaus are wide and well defined, the direct resistivity becomes zero inside the Hall plateaus and displays sharp peaks at the edges of the Hall plateaus. The Shubnikov-de Haas oscillations can be analyzed using semi-classical theoretical approaches and information about the geometry of the Fermi surface can be extracted.\cite{EtoPRB2010xy} This type of analysis has been intensely used in the search of the 2-dimensional component of the Fermi surface of 3-dimensional topological insulators, which can be related to the topological surface states. Besides the Shubnikov-de Haas oscillations, the widths of the Hall plateaus could be used to assess the degree of disorder present in the samples. Our hope is that the present technique will enable us to reverse engineer the experimental results on the resistivity tensor to a point where we can pin-point a particular microscopic model for a given sample.

The transport simulations in the presence magnetic fields present the same technical difficulties as before since the effect of the magnetic fields is introduced via the Peierls substitution as discussed. However, to resolve all the expected features, we will need to increase the simulation box. In this section we only want to demonstrate that we can indeed perform such simulations using the non-commutative formalism, more precisely, that we can resolve Hall plateaus and Shubnikov de Haas oscillations when the resistivity tensor is plotted as function of electron density. These features are easily reproduced when the data is plotted as function of the Fermi energy, but as we shall see, they disappear when plotted as function of electron density (which is how the data is plotted in the experiments), unless fairly strong disorder is present. As such, resolving the Hall plateaus in a transport simulation is a highly non-trivial problem, which even for the simple Integer Quantum Hall Effect has not been completely solved.  

\begin{figure}
\center
  \includegraphics[width=7cm]{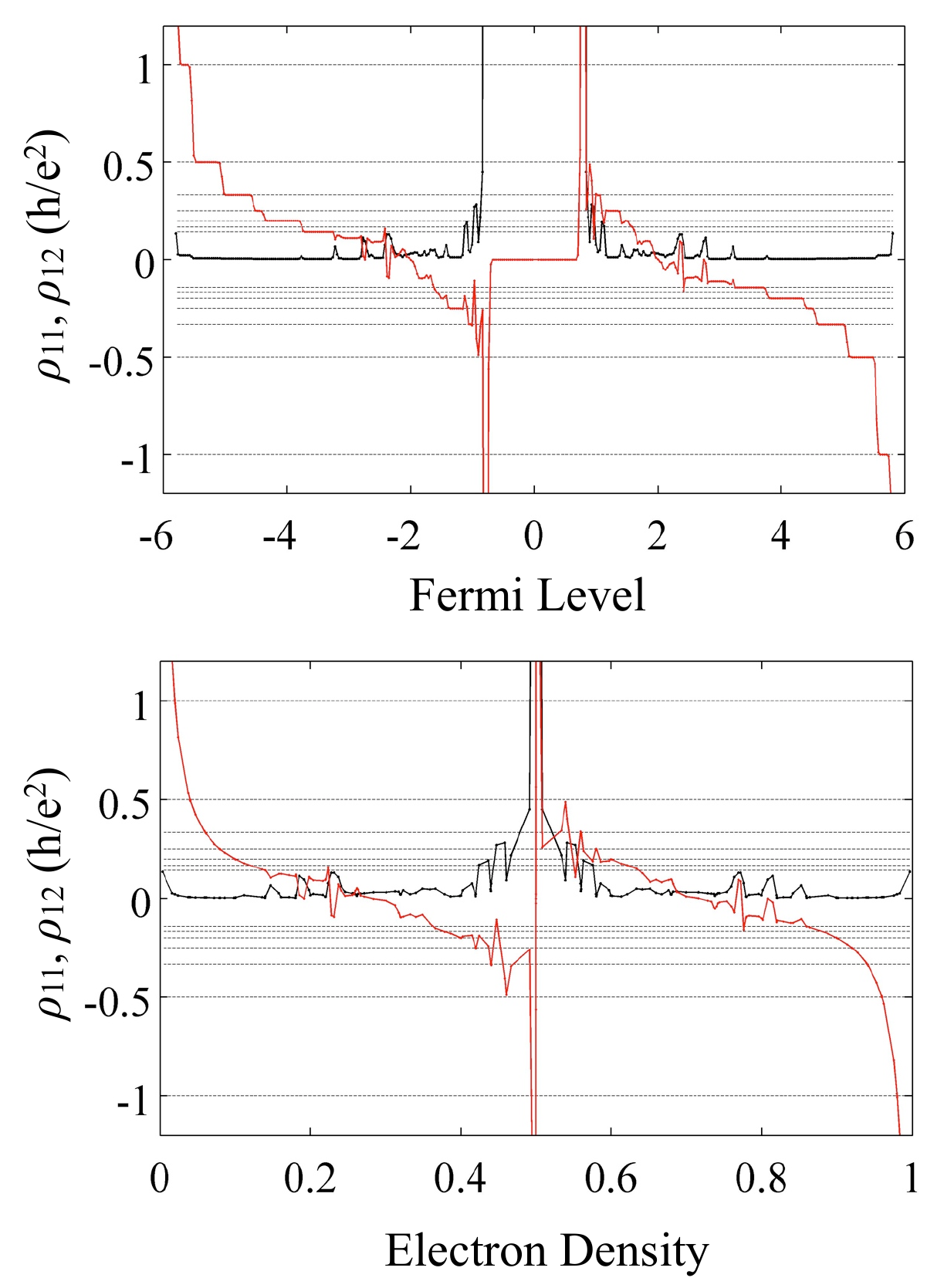}\\
  \caption{Upper panel: The diagonal and Hall resistivities for the topological case ($M=6$), computed at a fixed magnetic flux $\phi=0.08\times \phi_0$ while the Fermi energy was swept over the entire energy spectrum. The lattice size in these simulations was 50$\times$50, $W=0$ and $kT=1/\tau_{\mathrm{rel}}=0.01$. Lower panel: The data from the upper panel is replotted as function of the electron density.}
 \label{BCondW0}
\end{figure} 

We now start the discussion of our simulation results. The energy spectrum of periodic systems under magnetic fields have a Hofstadter fractal structure.\cite{HofstadterPRB1976km} Fig.~\ref{BDOSW0} shows the density of states (DOS) as function of energy and magnetic field, when the model is in the topological and trivial phases, with the disorder turned off. As it was already pointed out in Ref.~\onlinecite{Koenig:2008so}, there is an interesting qualitative difference in the behavior of the energy spectrum for the topological and trivial cases. What one basically sees in  Fig.~\ref{BDOSW0} is two Hofstadter butterflies, one below and one above the spectral gap, that are attracted to each other in the topological case, and are repelled by each other in the trivial case. The behavior of DOS with the magnetic field reminds us of the levitation and annihilation phenomenon discussed in the previous sections, but the cause of this behavior is not understood yet. Given that the insulating gap of most topological materials is small, this effect must be definitely taken into account when interpreting the experimental data.

\begin{figure}
\center
  \includegraphics[width=8.6 cm]{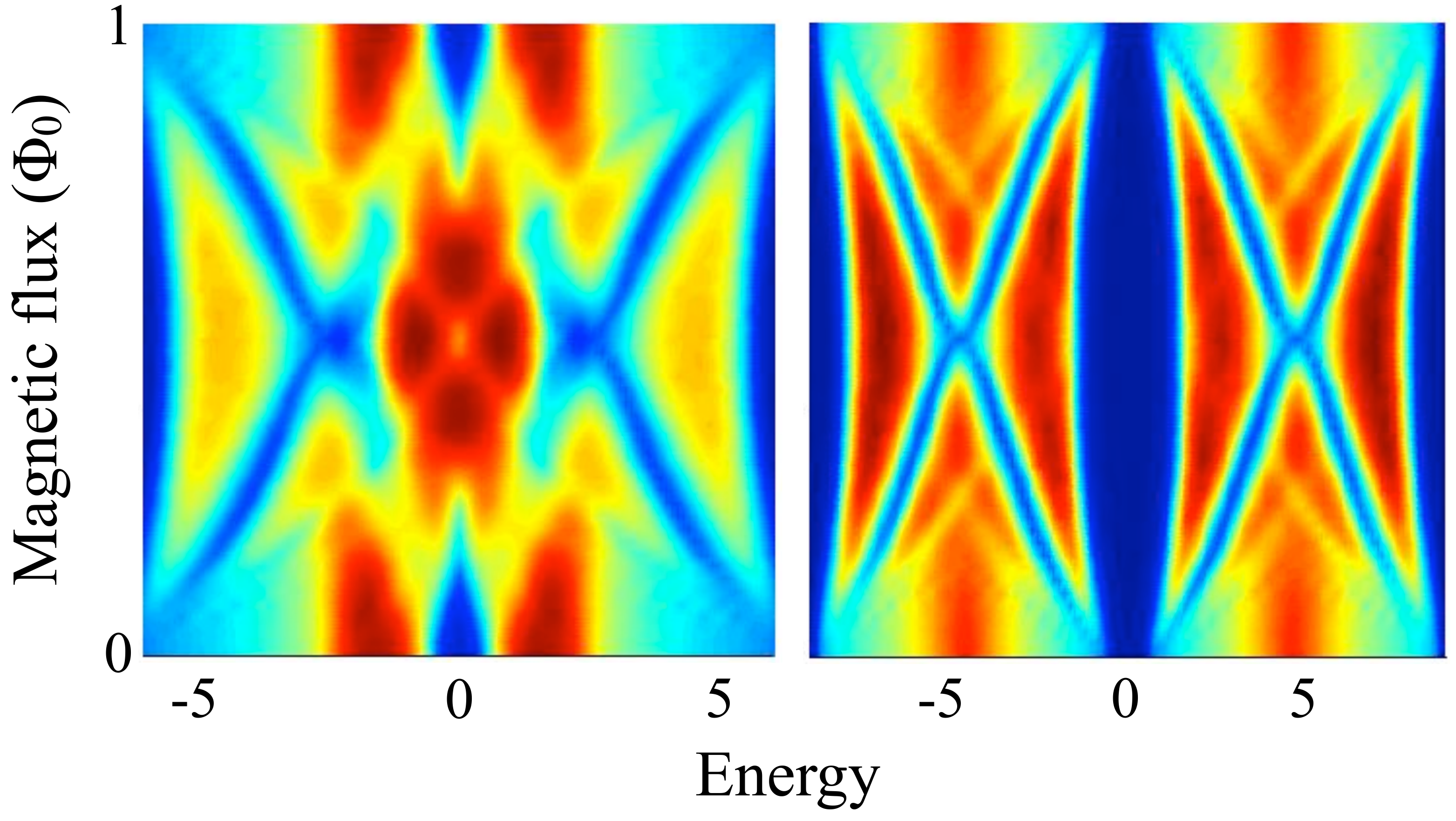}\\
  \caption{Same as Fig.~\ref{BDOSW0} but for $W=3$.}
 \label{BDOSW3}
\end{figure}

Now the Hofstadter spectrum is complex and has a fractal nature, but still at low magnetic fields one can distinguish clear thin Landau bands in Fig.~\ref{BDOSW0}. Of course, if we zoom in on these bands, we will see the whole structure repeating itself over and over again. But at the scale of Fig.~\ref{BDOSW0}, when the Fermi level is located in between these Landau bands, the Hall conductivity or resistivity should be quantized and the direct conductivity or resistivity should display a local minimum (the simulations are at finite temperature so we don't expect a strictly null direct conductivity). The Hall conductivity or resistivity should jump to the next Hall plateau when the Fermi level is brushed over a Landau band, while the direct conductivity or resistivity should have a spike. The Landau bands are clearly contoured at the bottom or at the top of the energy spectrum but, at the scale of Fig.~\ref{BDOSW0}, they are not well resolved at the edges of the insulating gap where one will be most interested in. As such, our transport simulations in which the Fermi level is varied over the entire energy spectrum and are done at the same scale as that of Fig.~\ref{BDOSW0}, are expected to show quantized Hall plateaus only at the bottom and at the top of the energy spectrum. More refined simulations which hopefully could resolve the energy region near the gap edges will be presented elsewhere. 

\begin{figure}
\center
  \includegraphics[width=7.1cm]{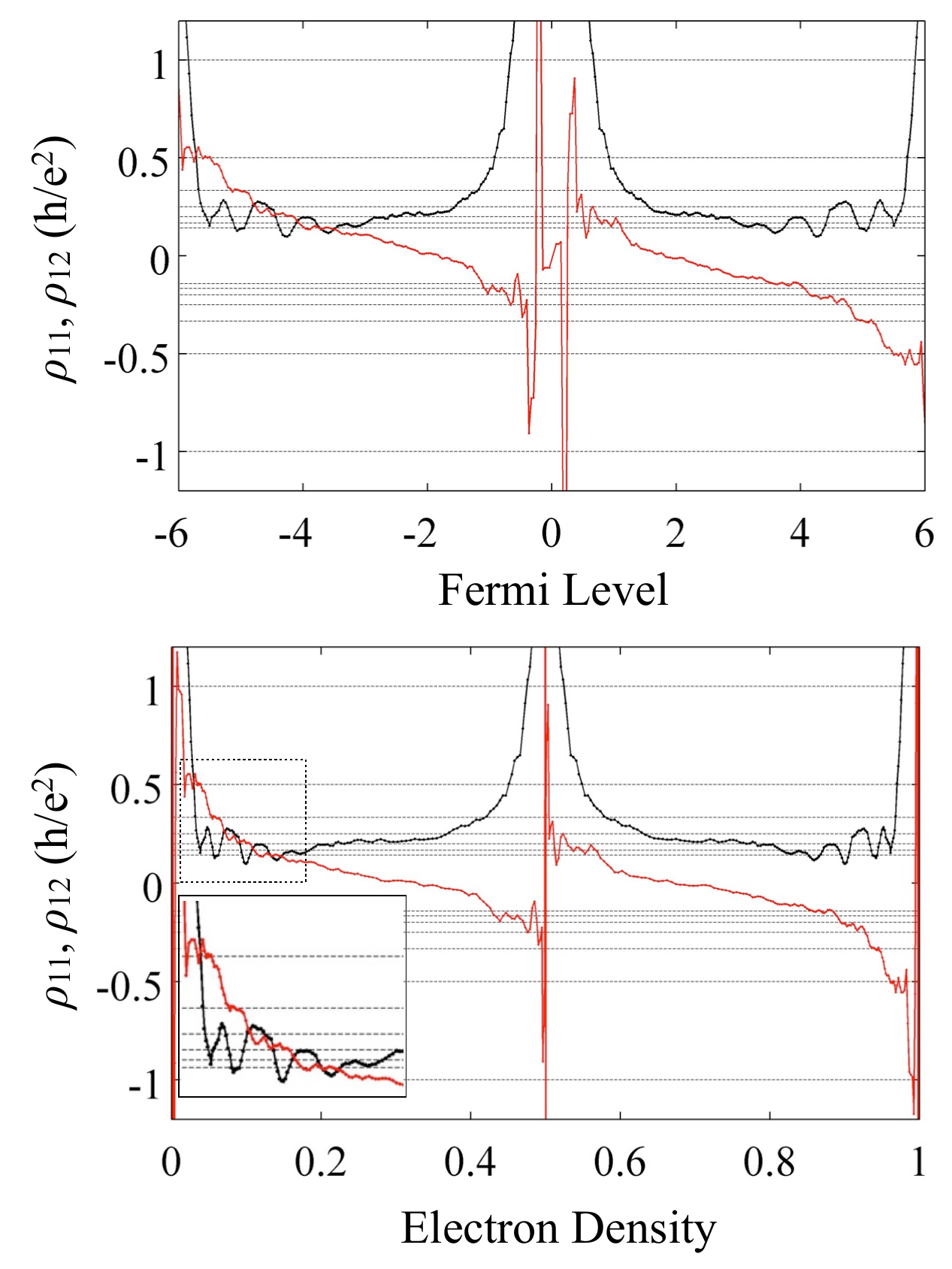}\\
  \caption{Same as Fig.~\ref{BCondW0} but for $W=3$.}
 \label{BCondW3}
\end{figure}

Fig.~\ref{BCondW0} reports the simulated direct and Hall resistivities for a magnetic flux per unit cell of $\phi=0.08$ $\phi_0$. The simulations were performed on a $50\times 50$ lattice. In Fig.~\ref{BCondW0}(a) the results are plotted as function of Fermi energy, while in Fig.~\ref{BCondW0}(b) the results are plotted as function of electron density. In the first case, the Hall plateaus appear very clearly contoured and the spikes in the direct resistivity can be seen whenever the sampled energies fell between two Hall plateaus. However, when the data is plotted as function of electron density, the Hall plateaus are reduced to a single point because the electron density does not vary when the Fermi level sweeps over the clean spectral gaps. As a result, the Hall plateaus disappear. When the disorder is turned on, the Hofstadter spectrum is smoothed out by the localized states that are pulled out the Landau bands, as one can see in Fig.~\ref{BDOSW3} where we plot the DOS for $W=3$. The results of the transport simulations for $W=3$ are reported in Fig.~\ref{BCondW3}. The magnetic flux per unit cell and the lattice size were fixed at the same values as for Fig.~\ref{BCondW0}. The plots in Fig.~\ref{BDOSW3} are obtained with a single disorder configuration. To read the data in Fig.~\ref{BDOSW3}, it is best to look first at the direct conductivity because it displays four clear dips. Examining the Hall resistivity plots, we see that they display a Hall plateau at each of these dips. The Hall plateaus are not perfectly quantized but their widths are clearly identifiable and that is enough for comparisons with the experiments. Very importantly, the widths of the Hall plateaus remain finite when the data is plotted as function of the electron density.

\section{Conclusions}

The main purpose of the present work was to introduce the reader to the noncommutative theory of charge transport and to present an efficient and fast converging numerical implementation of the noncommutative Kubo formula. As an interesting application, we chose a model of a topological insulator where we were able to map the diagonal conductivity as function of Fermi level, disorder strength and temperature. This enabled us to demonstrate that the topological phase is surrounded by a diffusive metallic phase where the transport coefficients increase as the temperature is lowered. This means that disorder alone can drive the system from a topological insulator into a diffusive metal. We postulate that this is what is actually happening in the current experimental observations where all the samples showed metallic bulk so far. When we introduce a perpendicular magnetic field, we found an intriguing behavior of the energy spectrum in that the valence and conduction bands move towards each other as the strength of the field is increased, until they touch and merge. In our transport simulations at fixed magnetic flux and in the presence of disorder, we were able to resolve a few Hall plateaus even when the conductivity was plotted as function of electron density.

\section*{Acknowledgments}

This work was supported by the U.S. NSF grants DMS-1066045 and DMR-1056168. 


%

\end{document}